\documentclass[a4paper,12pt]{article}
\pdfoutput=1
\usepackage{amssymb}
\usepackage{amsmath}
\usepackage{epsfig}
\usepackage{graphicx}
\usepackage{slashed}
\usepackage{color}
\usepackage{ulem}

\makeatletter
\renewcommand{\theequation}{\thesection.\arabic{equation}}
\@addtoreset{equation}{section}
\makeatother
\def\be{\begin{equation}}
\def\ee{\end{equation}}
\def\bea{\begin{eqnarray}}
\def\eea{\end{eqnarray}}
\def\eq{\begin{eqnarray}}
\def\eqx{\end{eqnarray}}

\def\({\left(}
\def\){\right)}
\def\<{\left<}
\def\>{\right>}

\def\be{\begin{equation}}
\def\ee{\end{equation}}
\def\ben{\begin{eqnarray}}
\def\een{\end{eqnarray}}
\def\({\left(}
\def\){\right)}
\def\<{\left<}
\def\>{\right>}
\def\!{\right|}
\def\|{\left|}

\def\[{\left[}
\def\]{\right]}

\def\+{\bar}

\def\tth{{\negthinspace}}
\def\tDelta{{\tilde{\Delta}}}

\usepackage{hyperref}
\hypersetup{colorlinks,%
  citecolor=blue,%
  linkcolor=blue,%
  pdftex}

\begin{document}

\begin{titlepage}
\vskip1cm
\begin{flushright}
\end{flushright}
\vskip0.25cm
\centerline{
\bf \large 
Transparentizing Black Holes to Eternal Traversable 
Wormholes 
} 
\vskip1cm \centerline{ \textsc{
 Dongsu Bak,$^{\, \tt a,c}$  Chanju Kim,$^{\, \tt b}$ Sang-Heon Yi$^{\, \tt a}$} }
\vspace{1cm} 
\centerline{\sl  a) Physics Department,
University of Seoul, Seoul 02504 \rm KOREA}
 \vskip0.3cm
 \centerline{\sl b) Department of Physics, Ewha Womans University,
  Seoul 03760 \rm KOREA}
   \vskip0.3cm
 \centerline{\sl c) Natural Science Research Institute,
University of Seoul, 
Seoul 02504 \rm KOREA}
 \vskip0.4cm
 \centerline{
\tt{(dsbak@uos.ac.kr,\,cjkim@ewha.ac.kr,\,shyi704@uos.ac.kr})} 
  \vspace{2cm}
\centerline{ABSTRACT} \vspace{0.75cm} 
{
\noindent 
We present the gravity description of evaporating black holes that end up 
with eternal traversable wormholes 
where every would-be behind horizon degree is available in asymptotic regions.  
The transition is explicitly realized by a time-dependent bulk solution in
the two-dimensional Einstein-dilaton gravity. In this solution, the initial 
AdS$_2$ black hole 
 is evolved into an eternal traversable wormhole
free of any singularity, which may be dubbed as
transparentization of black holes to eternal traversable wormholes.
The bulk construction completely matches with the boundary description 
governed by the Schwarzian boundary theory. We also obtain solutions
describing eternal traversable wormholes as well as excitations by an additional matter and graviton
oscillations on eternal traversable wormholes, which show that the eternal traversable wormhole states are gapped and non-chaotic.
Embedding the 2d solution into a 4d  traversable wormhole connecting two magnetically
charged holes, we discuss 4d scattering of a wave incident upon one
end of the traversable wormhole.
}

\end{titlepage}

%
\section{Introduction}
Recently, there are some interests in traversable wormholes in the context of the AdS/CFT correspondence~\cite{Gao:2016bin,Maldacena:2017axo,Maldacena:2018lmt,Bao:2018msr,Gao:2018yzk,Garcia-Garcia:2019poj}. Especially, AdS$_{2}$ wormholes have been explored in various contexts, since those have a dual description in the so-called SYK model~\cite{SYK} and could also be described via the  boundary Schwarzian theory~\cite{Maldacena:2016upp}.  (See~\cite{Sarosi:2017ykf,Rosenhaus:2018dtp} for a review.) In this regard, the construction of these traversable wormholes relies crucially on the non-local bulk interaction corresponding to the double trace deformation in the boundary theory. On the other hand,
AdS$_{2}$ space can be realized as the part of the near horizon geometry of extremal black holes and may be utilized as the model for higher dimensional black holes or wormholes. Along this line, traversable wormholes in asymptotically flat four-dimensional spacetime are constructed  in Einstein-Maxwell theory with charged massless fermions~\cite{Maldacena:2018gjk}. Though these traversable wormholes are unstable, those do not violate either the causality conditions or the (achronal) averaged null energy condition \cite{Anabalon:2018rzq}.  Since it has been regarded, for a long time, that traversable wormholes in Einstein gravity are difficult to construct  because of those conditions,  it would be of great interest to investigate further the possibility of traversable wormholes and their related physics. This might have some implications for various issues in black hole physics  such as  the information loss problem, which may also shed some lights  on the entanglement, the entropy~\cite{Israel:1967wq,Maldacena:2001kr,Maldacena:2013xja} and the complexity~\cite{Susskind:2017ney,Susskind:2018pmk} in the dual quantum system.

In this paper, we will continue to explore traversable AdS$_{2}$ wormholes from the bulk viewpoint in the two-dimensional Einstein-dilaton theory~\cite{Jackiw:1984je,Teitelboim:1983ux,Almheiri:2014cka}, and then consider four-dimensional traversable wormholes through our two-dimensional construction. As was constructed in Ref.~\cite{Maldacena:2018lmt}, it is possible to construct eternal traversable wormholes (ETWs)  in the 1d Schwarzian theory, which was shown to be the boundary reduction of the  2d bulk Einstein-dilaton theory~\cite{Maldacena:2016upp}. Though the equivalence of the boundary Schwarzian theory and its bulk Einstein-dilaton theory is derived in the context of the nearly AdS$_{2}$ gravity in Ref.~\cite{Maldacena:2016upp}, it would be better to have a direct bulk description of traversable AdS$_{2}$ wormholes in the 2d Einstein-dilaton theory. This bulk  gravity description seems more desirable in regard to various issues in  higher dimensional black hole/wormhole physics.  For instance, the behavior of singularity may be traced more or less directly in this bulk description. And the bulk causal structure is more manifest in our bulk description. These aspects justify partially why we need  to study the bulk description in some detail.

In our model, we show that  it is possible to construct time-dependent objects which were black holes with past singularities and then  become non-singular traversable wormholes in the end.  This is a simple concrete example, which shows how the would-be singularity 
is removed in a controlled way.  And then, we construct static solutions describing eternal wormholes, which  correspond to the simplest solutions in the Schwarzian theory. It turns out that their description in the bulk are a bit complicated and have noticeable features, which also  indicate the necessity of our exploration into the detailed bulk construction. Furthermore, through small perturbations on our bulk solutions, one can see that  the seemingly thermalized and scrambled information could be recovered in the end.  

This paper is organized as follows. In section 2, we summarize briefly our previous work~\cite{Bak:2018txn}, to give our notations and to present our setups. In section 3, we present our results on the bulk dilaton field deformation by the double trace deformation interaction between two boundary sources. This section is a bit technical and so some computational details are relegated to Appendices. In section 4, we present the solutions showing the transition from black holes to eternal traversable wormholes. Using these solutions, one can see that the singularity inside black holes disappears in a controlled way. And then, we point out that the intrinsic ambiguity in our construction could be remedied by an appropriate physical consideration. Some aspects on  sending signals from one side to the other one in the two-sided spacetime is also discussed. In section 5, we present our bulk construction of eternal traversable wormholes dual to those in the Schwarzian theory. We also give bulk solutions describing the transition from eternal traversable wormholes into black holes. We show the complete matching between the boundary and bulk results.  In section 6, we add a single matter excitation in the bulk, and see its effects on the eternal traversable wormholes. It turns out that this bulk excitation also matches completely with the boundary Schwarzian one and reveals that the traversable wormhole state is distinguished from the black hole one. In section 7, we consider the boundary graviton oscillation on the eternal traversable wormholes, which is described by 
the double-trace interaction with a time-dependent coefficient. 
It is checked that the boundary and bulk description are completely equivalent and is shown that the eternal traversable wormholes are gapped, indeed.  In section 8, four-dimensional long traversable wormholes are discussed in the context of the scattering cross section. We summarize our results and discuss some open questions in the final section.

\section{Boundary description and Bulk perturbation
}\label{sec2}

In this section we briefly review our previous work \cite{Bak:2018txn} to fix notations
and generalize the system to incorporate eternal traversable wormholes.
We begin with the two-dimensional Einstein-dilaton gravity coupled to a scalar field $\chi$ with mass $m$,
\begin{equation}
I=I_\textrm{top}+\frac1{16\pi G}\int_M d^2 x \sqrt{-g}\, 
\phi \left( R+\frac{2}{\ell^2}\right) + I_\textrm{surf} + I_M(g, \chi)\,,
\end{equation}
where 
\begin{align}
I_\textrm{top}&= \frac{\phi_0}{16\pi G}\int_M d^2 x \sqrt{-g}\, R\,,  \notag \\
I_\textrm{surf} &= -\frac1{8\pi G}\int_{\partial M} 
	             \sqrt{-\gamma}\, (\phi_0 + \phi ) \, K \,, \notag \\
I_M &= -\frac12 \int_M d^2 x \sqrt{-g} ( \nabla \chi \cdot \nabla \chi + m^2
\chi^2 ) \,,
\end{align}
and $\gamma_{ij}$ and $K$ denote the induced metric and  
the extrinsic curvature at $\partial M$, respectively.

The equation of motion for the dilaton field $\phi$ sets the metric to be
AdS$_2$ which can be written in the global coordinates as
\begin{equation}
ds^2 =\frac{\ell^2}{\cos^2 \mu} \left(-d\tau^2 + d\mu^2  \right)\,,
\end{equation}
where $\mu$ is ranged over $[-\frac{\pi}{2},\frac{\pi}{2}]$.
The equations of motion for the metric read
\begin{equation} \label{eqphi}
\nabla_a \nabla_b \phi -g_{ab} \nabla^2 \phi + g_{ab} \phi = -8 \pi G T_{ab}\,,
\end{equation}
where $T_{ab}$ is the stress tensor of the scalar field,
\begin{equation}
T_{ab} = \nabla_a \chi \nabla_b \chi  -\frac{1}{2} g_{ab} 
        \left( \nabla \chi \cdot \nabla \chi + m^2 \chi^2 \right).
\end{equation}
In this note, we shall set $8\pi G =1$ for the simplicity of our presentation and recover it whenever the Newton constant  $G$ is necessary.
With $T_{ab}=0$, the general vacuum solution for the dilaton field 
can be obtained in the form \cite{Bak:2018txn},
\begin{equation} \label{dilaton}
\phi= \phi_{bh}(L,b,\tau_{B})\equiv \bar\phi L
 \frac{(b+b^{-1}) \cos(\tau-\tau_B) -(b-b^{-1}) \sin \mu}{2 \cos \mu} \,,
\end{equation}
where we choose $b \ge 0$. By the coordinate transformation 
\begin{align} 
r &= \frac\phi{\bar\phi}= L \frac{(b+b^{-1}) \cos(\tau-\tau_B) -(b-b^{-1}) 
                              \sin \mu}{2 \cos \mu} \,,\notag \\
 \tanh \frac{t L }{\ell^2} &=\frac{2\sin (\tau-\tau_B)}{(b+b^{-1}) \sin \mu -(b-b^{-1}) \cos (\tau-\tau_B)}\,,
 \label{coorb}
\end{align}
one is led to the corresponding AdS$_{2}$ black hole metric 
\begin{equation} \label{btz}
ds^2= - \frac{r^2-L^2}{\ell^2} dt^2+ \frac{\ell^2}{r^2-L^2} dr^2\,.
\end{equation}

Boundary values of the metric and the dilaton are fixed in $\epsilon
\rightarrow 0$ limit as
\begin{equation} \label{bcondition}
	ds^2|_{\partial M} = -\frac1{\epsilon^2}d\tilde{u}^2\,, \qquad
	\phi|_{\partial M} = \frac{\ell \, \bar\phi }\epsilon\,,
\end{equation}
where $\tilde{u}$ denotes the boundary time.
For the bulk scalar field $\chi$, we shall impose the boundary condition
corresponding to the double trace deformation of the boundary theory \cite{Gao:2016bin},
\begin{equation} \label{defham}
\delta H(\tilde u) = -h(\tilde u){\cal O}_{R}(\tilde u){\cal O}_{L}(\tilde u)\,,
\end{equation}
where the subscript $L/R$ refers to the left/right boundary and
$\mathcal{O}_{R,L}$ are scalar operators of dimension $\Delta$ dual to
$\chi$. Then the coupling $h(\tilde u)$ has dimension $1-2\Delta$.%
\footnote{For the double trace deformation, we shall consider the case $0<\Delta<\frac12$ so that 
the deformation \eqref{defham} is relevant. This is possible when the mass
is in the range $-1/4 < m^2 < 0$ as
$ \Delta = \frac{1}{2} \left( 1 - \sqrt{1+4m^2}\right) $.}

Since the metric is not affected by the presence of the matter field,
the back reaction due to the deformation \eqref{defham} is completely
described by the dynamics of the dilaton field $\phi$ which is governed by \eqref{eqphi}.
The general solution of \eqref{eqphi} can be written as \cite{Bak:2018txn}
\begin{equation}
	\phi = \bar\phi L \frac{\cos\tau}{\cos\mu} + \varphi,
\end{equation}
where
\begin{equation} \label{DilDef}
\varphi(u,v) 
= \int^{u}_{u_{0}}dp \frac{\sin(p-u)\cos(p-v)}{\cos(u-v)} T_{uu}(p,v),
\end{equation}
upto homogeneous terms that may be fixed in various ways and $u$, $v$ are global null coordinates defined by
\begin{equation}
u = \frac12(\tau + \mu),\qquad
v = \frac12(\tau - \mu).
\end{equation}
In the case that the coupling $h(\tilde u)$ is nonzero only within the 
Rindler wedge of the metric \eqref{btz},
an explicit bulk solution was obtained in \cite{Bak:2018txn} by computing the 1-loop 
stress tensor for the deformation \eqref{defham}. In this paper, 
we would like to consider a general case  that the deformation may
persist indefinitely.

The leading correction to the bulk two-point function is
\begin{align}
\delta \left\langle \frac12\{\chi(x),\chi(x')\} \right\rangle
&= \frac{i}2 \int_{\tau_i}^\tau d\tau_s 
 \langle \{ [ \delta H(\tau_s), \chi(x) ], \chi(x') \} \rangle
 + (x \leftrightarrow x') \notag \\
 &\equiv F(x|x') + F(x'|x),
\end{align}
where $\tau_i$ is the time that the double trace deformation is turned on.


One may equivalently describe the back reaction of the deformation
by considering the boundary effective action which consists of
the Schwarzian derivatives at boundaries and an interaction
term~\cite{Maldacena:2017axo,Maldacena:2018lmt},
\begin{equation} \label{sbdry}
S = \int d\tilde{u} \bigg[ -\phi_{l}\Big\{ \tan \frac{\tau_{l}(\tilde{u})}{2},\tilde{u}\Big\}  -\phi_{r}\Big\{ \tan \frac{\tau_{r}(\tilde{u})}{2},\tilde{u} \Big\}  + \frac{g}{2^{2\Delta}}\bigg(\frac{\tau'_{l}(\tilde{u})\tau'_{r}(\tilde{u})}{\cos^{2}\frac{\tau_{l}(\tilde{u})-\tau_{r}(\tilde{u})}{2}} \bigg)^\Delta\bigg]\,,
\end{equation}
where $\phi_{l}=\phi_{r}$ can be identified with $\bar{\phi}$ in the bulk
and $g$ is a coupling proportional to the bulk parameter $h$.
By comparing the solution of the equation of motion of \eqref{sbdry}
and the bulk solution \eqref{DefDil} with the boundary condition
\eqref{bcondition}, one can identify $g$ as \cite{Bak:2018txn}
\begin{equation}  \label{gtoh}
g = \frac{h}{2\pi}\frac{2^{2\Delta-1}\Gamma^{2}(\Delta)}{\Gamma(2\Delta)}.
\end{equation}
Then, to the leading order in $g$, the boundary dynamics completely matches 
with the bulk solution.



\section{Dilaton Field}\label{sec3}
In this section we shall present the back-reacted dilaton field solution by the non-local bulk interaction corresponding to the double trace boundary deformation.  First, let us recall that the bulk to boundary two point functions are given by
\begin{align}    \label{}
K_{L}(\tau- \tau_s,\mu)&  =    \frac{2^{\Delta-2}\Gamma^{2}(\Delta)}{\pi \Gamma(2\Delta)} \Big[\frac{\cos\mu}{\cos(\tau-\tau_{s}) + \sin \mu} \Big]^{\Delta}\,, \nonumber \\
K_{R}(\tau- \tau_s,\mu)&=  \frac{2^{\Delta-2}\Gamma^{2}(\Delta)}{\pi \Gamma(2\Delta)} \Big[ \frac{\cos\mu}{\cos(\tau-\tau_{s}) - \sin \mu} \Big]^{\Delta} \,, 
\end{align}
whose normalization is chosen from the bulk two point functions as shown in Appendix~\ref{AppA}. 
Here, $L/R$  represent that the locations of the double trace interaction at each boundaries, which   are  taken, in global $(\tau,\mu)$ coordinates, as $(\tau_{s}, - \frac{\pi}{2})$ for  $K_{L}$  and as $(\tau_{s},  \frac{\pi}{2})$ for $K_{R}$, respectively. Since the exponent $\Delta$ is not an integer in our case, the phase of $K_{L/R}$ functions should be chosen appropriately, whenever the values inside brackets are negative. While this phase is naturally  chosen via the $i\epsilon$-prescription in the Rindler wedges in~\cite{Bak:2018txn,Gao:2016bin}, the phase assignments in the case of eternal traversable wormholes  is a bit more involved and so those are relegated to Appendix~\ref{AppA}.

By following the procedure in~\cite{Bak:2018txn}, one can see that the deformation of bulk (Hadamard) two-point  function   by  the double trace interaction   can  be written as, $ \delta \langle \frac{1}{2}\{\chi(x),\chi(x')\} \rangle = F(x|x') + F(x'|x)$.   Before going ahead, one may note that the boundary Hamiltonian deformation in Eq.~(\ref{defham})  can be rewritten in terms of the bulk global time $\tau$  as
\begin{equation} \label{}
\delta H(\tau) = -h(\tau)  {\textstyle (\frac{d\tau}{d\tilde{u}})^{2\Delta-1} }  {\cal O}_{R}(\tau){\cal O}_{L}(\tau)\,,
\end{equation}
where we have used the conformal transformation ${\cal O}(\tau) = (\frac{d\tau}{d\tilde{u}})^{\Delta} {\cal O}(\tilde{u})$ and have rewritten $h(\tilde{u})$ in terms of $\tau$. We will also set $h(\tau) = h = const.$  for  the simplicity.  Then,  following the procedure in~\cite{Gao:2016bin,Bak:2018txn}, one can see that  the  function $F(x|x')$ is taken as\footnote{In the following, we will consider the case $\tau_{f} > \tau$ only, since we are interested in the limit $\tau_{f}\rightarrow \infty$. } 
\begin{equation} \label{Fxx}
F(x|x')  =  \frac{i h }{2}\int^{\tau_{f}}_{\tau_{i}} d\tau_{s}  {\textstyle (\frac{d\tau_{s}}{d\tilde{u}})^{2\Delta-1} }  \Big[ {\scriptstyle   K^{+}_{L}(\tau'-\tau_{s},\mu')K^{-}_{R}(\tau-\tau_{s},\mu) + K^{+}_{R}(\tau'-\tau_{s},\mu) K^{-}_{L}(\tau-\tau_{s},\mu)   } \Big] \theta(\tau-\tau_{s}) \,,
\end{equation}
where $\theta(x)$ denotes the unit step function and  $K^{\pm}_{L/R}$'s are defined by
\begin{equation} \label{}
K^{\pm}_{L/R} (\tau,\mu) \equiv K_{L/R}(\tau+i\epsilon,\mu) \pm K_{L/R}(\tau-i\epsilon, \mu) \Big|_{\epsilon\rightarrow 0}\,.
\end{equation}
In the following, we have interested in the case of  constant $\frac{d\tau}{d\tilde{u}}$, which will be related to the eternal traversable wormholes. 
Then, it is convenient to   introduce  a new constant parameter $\bar{h}$ as  
\begin{equation} \label{barh}
\bar{h} \equiv h \Big(\frac{d\tau_{s}}{d\tilde{u}}\Big)^{2\Delta-1} = const.
\end{equation}
 A more generic case will be discussed briefly in a later section.

\begin{figure}[th]   
\begin{center}
\includegraphics[width=9cm 
]{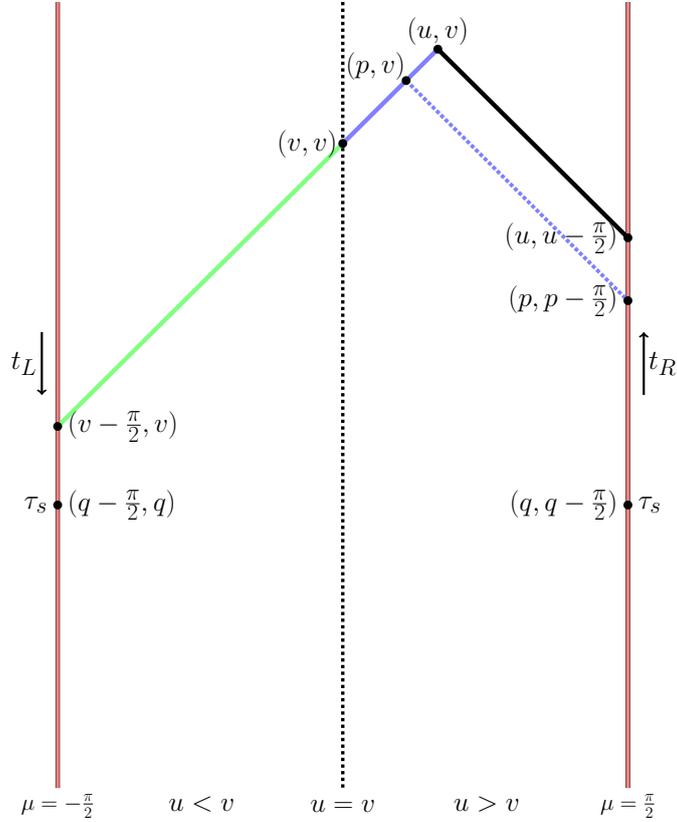}
\caption{The evaluation point for $\varphi$ and its related ones  written in $(u,v)$ coordinates in the global AdS, which are defined by  $\tau=u+v$, $\mu=u-v$. }
\label{UVdia}
\end{center}
\end{figure}

One may note that $K^{-}_{L/R}(\tau,\mu)\theta(\tau)$  are  nothing but the retarded two point functions and  vanish between the space-like separated points. Note also that $\tau_{i}/\tau_{f}$'s are  the turning-on/off times of the double trace interaction.  According to our phase assignments in Appendix~\ref{AppA}, one can set 
\begin{align}    \label{phaseF}
K^{+}_{L/R} (\tau,\mu) &= 2\cos \nu_{L/R}~ |K_{L/R} (\tau,\mu)|\,,   \\
 K^{-}_{L/R} (\tau,\mu) &= -2i\sin \nu_{L/R}~ |K_{L/R} (\tau,\mu) |~ \theta(-d_{L/R})\,, \nonumber
\end{align}
where we have  introduced $d_{L/R}$ functions as\footnote{The last $\theta$-function is incorporated in consistency with the usual $i\epsilon$-prescription  in Ref.~\cite{Gao:2016bin,Bak:2018txn}.}
\begin{equation} \label{dfunc}
d_{L/R}(\tau-\tau_{s}, \mu) = \cos(\tau-\tau_{s}) \pm  \sin\mu\,.
\end{equation}
It turns out that these functions, $d_{L/R}(\tau-\tau_{s}, \mu)$ are responsible for  the phase assignments and the nature of causal relations between the evaluation point and the source points.  
%
%
Note that the locations of the double trace  interaction or the left/right sources can be written,  in terms of a new coordinate $q$  (see Figure~\ref{UVdia}), as 
\begin{equation} \label{qvar}
\tau_{s} = 2q-\frac{\pi}{2}\,,  \qquad \mu = \mp \frac{\pi}{2}\,,
\end{equation}
where $q$ may be regarded as the $u$-coordinate for the location of the right boundary source  or as the  $v$-coordinate for that of the left boundary one.  Using this $q$ coordinate, one can see that $d_{L/R}(u,v\,;\,q)$ functions become
\begin{equation} \label{}
d_{L}(u,v\,;\,q) = 2\sin(q-v)\cos(q-u)\,,  \qquad d_{R} (u,v\,;\,q)= 2\cos(q-v)\sin(q-u)\,.
\end{equation}
From these functions, we can readily determine the  causal relation between the evaluation point $(u,v)$ and the source points $(\tau_{s},\mp\frac{\pi}{2})$.  For instance, the evaluation point $(u,v)$ is time-like separated from the left/right boundary sources  if and only if $d_{L/R} (u,v\,;\,q) <0 $.  In terms of $d_{L/R}(u,v\,;\,q)$ functions,  the function $F$ is given by
\begin{align}    \label{}
F(u,v|u',v')   & =   2\bar{h}\sin(\nu_{L}+\nu_{R})\bar{N}_{\Delta} \int^{q_{f}}_{q_{i}} dq \bigg[ {\textstyle  \Big(\frac{\cos(u'-v')}{|d_{L}(u',v'\,;\,q)|}\Big)^{\Delta} \Big(\frac{\cos(u-v)}{|d_{R}(u,v\,;\,q)|}\Big)^{\Delta}  {\scriptstyle \theta(-d_{R} )  }  }   \qquad  \qquad  \qquad \nonumber   \\  
& \qquad \qquad  \qquad  \qquad  \qquad \qquad  +     \Big\{ (u,v)  \leftrightarrow (u',v') \Big\}   {\scriptstyle \theta(-d_{L} )  }  
 \bigg] {\scriptstyle \theta\big(u+v - 2q+\frac{\pi}{2}\big) } \,, 
\end{align}
%
%
where $\bar{N}_{\Delta} \equiv 2\big[\frac{2^{\Delta-2}\Gamma^{2}(\Delta)}{\pi\Gamma(2\Delta)}\big]^{2}=  \frac{N_{\Delta} }{\sin\pi\Delta}$ and $q_{i/f} = \frac{1}{2}(\tau_{i/f}+\frac{\pi}{2})$. It is useful to observe that  the upper limit of $q$-integration reduces to $q_{\tau}\equiv \frac{1}{2}(\tau+\frac{\pi}{2})$ because of the $\theta$-function on $q$. 
By using the result in Eq.~(\ref{DilDef}),  the deformation of the dilaton field by the double trace interaction is shown to be given  by  (see Appendix~\ref{appb} for some details)
\begin{align}    \label{DilatonM}
\varphi(u,v)  & =  -2 \int^{u}_{u_{0}}dp \int^{q_{\tau}}_{q_{i}}dq    {\textstyle \bigg[   h_{L} \Big\{ 1+  {\scriptstyle (\Delta+1)}\frac{\sin(u-p)\sin(q-v)}{\cos(u-v)\cos(q-p)}  \Big\}  \theta(-d_{R} )   }  \nonumber \\ 
&\qquad \qquad \qquad \qquad \qquad   {\textstyle  + h_{R} \Big\{ 1- {\scriptstyle (\Delta+1)} \frac{\sin(u-p)\cos(q-v)}{\cos(u-v)\sin(q-p)}    \Big\} \theta(-d_{L} )   \bigg]  }  \,,
\end{align}
where $d_{L/R}\equiv d_{L/R}(p,v\,;\, q)$ and $h_{L/R}$  are given by  
\begin{align}    \label{hl}
h_{L}(p,v\,;\, q)  &= 2\bar{h}\sin(\nu_{L}+\nu_{R})\bar{N}_{\Delta}~  \partial_{u'}{\textstyle  \Big(\frac{\cos(u'-v')}{|d_{L}(u',v'\,;\,q)|}\Big)^{\Delta} \Big(\frac{\cos(p-v)}{|d_{R}(p,v\,;\,q)|}\Big)^{\Delta}   } \bigg|_{(u',v') \rightarrow (p,v)}\,,   \\
h_{R}(p,v\,;\, q) &=  2\bar{h}\sin(\nu_{L}+\nu_{R})\bar{N}_{\Delta} {\textstyle \Big(\frac{\cos(p-v)}{|d_{L}(p,v\,;\,q)|}\Big)^{\Delta}\partial_{u'}\Big(\frac{\cos(u'-v')}{|d_{R}(u',v'\,;\,q)|}\Big)^{\Delta} }\bigg|_{(u',v') \rightarrow (p,v)}\,.   \label{hr}
\end{align}

Some comments are in order. First, the point $(u,v)$ in Figure~\ref{UVdia} denotes the evaluation point for the dilaton deformation $\varphi$, while $(p,v)$ denotes the evaluation point for stress tensor $T_{ab}(p,v)$. Second, note that one may take $u_{0} = v - \frac{\pi}{2}$ in Eq.~(\ref{DilDef}) or in Eq.~(\ref{DilatonM}),  because $T_{uu}(p,v)$ cannot exist on $ p < v-\frac{\pi}{2}$. 
Third, because of the $\theta(-d_{L/R})$ functions, there would be no contribution  to the dilaton deformation $\varphi$,  whenever the evaluation point $(p,v)$ for $T_{uu}(p,v)$ is space-like separated  from the left/right boundary sources, respectively.  Fourth, the final expression of dilaton deformation $\varphi$ should have a $\pi$-periodicity anticipated from the physical consideration. 
Finally, in order to perform the integration in the dilaton field deformation analytically, it is  convenient to  swap the integration order in $p$ and $q$.

\setcounter{table}{0}
\renewcommand{\thetable}{\color{blue}{\bf A}}
\begin{table} 
\begin{center}
\begin{tabular}{|c|c|c|c|c|}
 \hline
 $q$-range &   ${\scriptstyle v< q < p }$  &  ${\scriptstyle p-\frac{\pi}{2} < q < v  }$  &  ${\scriptstyle v-\frac{\pi}{2} < q < p -\frac{\pi}{2} }$  & ${\scriptstyle   p-\pi <  q < v-\frac{\pi}{2}  }$ \\
\hline
 $d_{L}(p,v\,;\, q) $    &  $+$   &   $-$   &   $+$   &   $+$      \\
 \hline
 $d_{R}(p,v\,;\, q)$  &   $-$ &      $-$     &    $-$     &      $+$   \\
 \hline 
\end{tabular}
\end{center}
\caption{\color{black} $v <  p < u$ }
\label{Tab1}
%
\renewcommand{\thetable}{\color{blue}{\bf B}}
\begin{center}
\begin{tabular}{|c|c|c|c|c|}
 \hline
 $q$-range &  ${\scriptstyle p < q < v } $ &  ${\scriptstyle v-\frac{\pi}{2} < q < p } $  &  $ {\scriptstyle p-\frac{\pi}{2} < q < v -\frac{\pi}{2} }  $  &  ${\scriptstyle v-\pi <  q < p-\frac{\pi}{2} } $ \\
\hline
 $d_{L}(p,v\,;\, q)$    &  $-$   &   $-$   &   $-$   &   $+$      \\
 \hline
 $d_{R}(p,v\,;\, q)$  &   $+$ &      $-$     &    $+$     &      $+$   \\
 \hline 
\end{tabular}
\end{center}
\caption{\color{black} $v-\frac{\pi}{2} <  p <  v$ }
\label{Tab2}
\end{table}
\begin{figure}[htbp]   
\begin{center}
\includegraphics[width=0.45\textwidth]{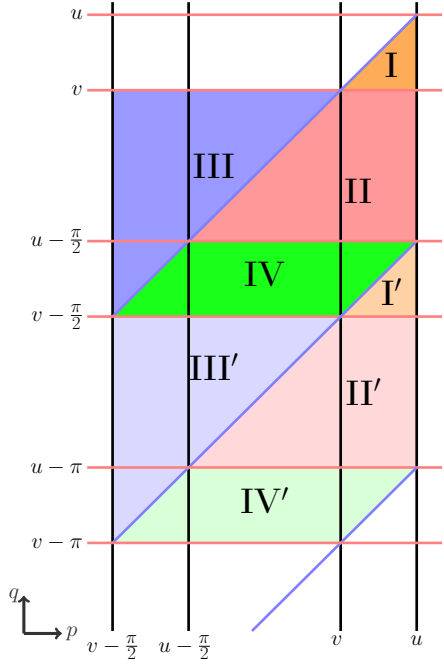}
\caption{The integration range is depicted in this diagram.}
\label{pqrange}
\end{center}
\end{figure} 
For definiteness, let us take the evaluation point $(u,v)$ of the dilaton deformation $\varphi$  satisfying the condition   $ u >  v$, which does not lose any generality\footnote{The special case of $u=v$ will be discussed  in conjunction with the extra contribution to $T_{ab}$ in the next section. }.  Under the assumption that $u > v$,  in order to  interchange the order of integration in $p$ and $q$, we need to classify the integration ranges of $p$ and $q$  as in the tables~\ref{Tab1} and \ref{Tab2}, where $+$ represents the space-like separation  ({\it i.e.} $d_{L/R} >0$) and $-$ does the time-like separation ({\it i.e.} $d_{L/R} < 0$)   between the evaluation point $(p,v)$ and the locations of source denoted by $q$. With this causality consideration, the classification of the integration regions can be depicted as in Figure~\ref{pqrange}. 
Let us focus on the {\bf I}, {\bf II}, {\bf III} and {\bf IV}  regions  in Figure~\ref{pqrange} or the left two columns in the tables~\ref{Tab1} and \ref{Tab2}, since the ${\bf I}'$, ${\bf II}'$, ${\bf III}'$ and ${\bf IV}'$ region turns out to be evaluated through a simple shift of variables.  
According to each {\bf I}, {\bf II}, {\bf III} and {\bf IV}  region in Figure~\ref{pqrange}, the integration order of $p$ and $q$ may be interchanged in each region  as 
\begin{align}    \label{Rearrange}
\int^{u}_{u_{0}}dp \int^{q_{\tau}}_{q_{i}}dq   = 
\left\{ \begin{array}{ll}    
\int^{u}_{q_{i}}dq\int^{u}_{q}dp    &  \quad {\rm for} \quad {\bf I}\,,   \\  
\int^{v}_{q_{i}}dq\int^{u}_{q}dp    &  \quad {\rm for} \quad   {\bf II}\,,   \\       
\int^{v}_{q_{i}}dq\int^{q}_{v-\frac{\pi}{2}}dp    &  \quad {\rm for} \quad   {\bf III}\,,   \\   
\int^{u-\frac{\pi}{2}}_{q_{i}}dq\int^{q+\frac{\pi}{2}}_{q}dp    &  \quad {\rm for} \quad   {\bf IV}\,,   
\end{array}  \right.   
\end{align}
where the equality means that the left hand side integration  region is divided into smaller regions  {\bf I}, {\bf II}, {\bf III} and {\bf IV} and needs to be added up according to the value of $q_{i}$.  
In this rearrangement of the integration order, one may note that the upper limit $q_{\tau}$ of the $q$-integration is always larger than $u$ and $v$, since $q$ variable is related to the location of sources and its final position  will be taken large indefinitely in our setup. Therefore, the upper limit of the $q$-integration reduces further to an appropriate value by the restriction given in the tables~\ref{Tab1} and~\ref{Tab2}.  The lower limit of the $p$-integration could also be determined in this way in each region. All these consideration is succinctly summarized in Figure~\ref{pqrange}. As is clear from Figure~\ref{pqrange}, one may note that the minimum value of $q_{i}$  in the $q$-integral is given by
\begin{align}    \label{Qmin}
q_{i}   = 
\left\{ \begin{array}{ll}    
v  &  \quad {\rm for} \quad {\bf I}\,,   \\  
u-\frac{\pi}{2}     &  \quad {\rm for} \quad   {\bf II}\,,   \\       
v-\frac{\pi}{2}   &  \quad {\rm for} \quad   {\bf III}  ~~{\rm and}~~  {\bf IV}\,.
\end{array}  \right.   
\end{align}
Since $\tau_{i} = 2q_{i}- \frac{\pi}{2}$ denotes the time when the double trace interaction is turned on,  it needs to be massaged in an appropriate way for the eternal wormholes which do not allow any fixed value for $\tau_{i}$  (see section~\ref{sec5} for this point).

We would like to emphasize that all other regions could be obtained by using the formulae in Appendix~\ref{appb}.   Especially, the computation in ${\bf I}'$, ${\bf II}'$, ${\bf III}'$ and ${\bf IV}'$ regions are completely parallel to the unprimed regions and easily obtained by the variable shift, as was shown  in Appendix~\ref{appb}. When $q_{i}$ is located beyond  the top of  {\bf I}, {\bf II}, {\bf III} and {\bf IV} regions ({\it i.e.} $q_{i} > u$), there is no contribution to the dilaton deformation because of causality. Beyond the bottom of ${\bf I}'$, ${\bf II}'$, ${\bf III}'$ and ${\bf IV}'$ regions  ({\it i.e.} $q_{i} < v -\pi$), we don't need further consideration because of the $\pi$-periodicity of the dilaton deformation $\varphi$ in the variable $q$.  That is to say, the same expression of $\varphi$ would be repeated below.

Before going ahead, it is useful to note that  $F(x | x')$ vanishes whenever  $\nu_{L}+\nu_{R} =0$, as can be seen from Eq.~(\ref{Fxx}) and Eq.~(\ref{phaseF}). This  is the case on the regions {\bf II},  ${\bf II}'$, {\bf IV} and ${\bf IV}'$, as shown in Eq.~(\ref{phasesum}).  Therefore, there is no contribution to dilaton deformation,  $\varphi$ from those regions.  Some computational details are relegated to Appendix~\ref{appb}. Here, we just summarize the results in the relevant region {\bf I},  {\bf III},  ${\bf I}'$ and ${\bf III}'$ only.

\underline{Region {\bf I}}:  As is clear from Figure~\ref{pqrange},  $q_{i} < u$ in this region.  When $q_{i} > v$, the contribution to the dilaton field deformation from the region {\bf I} is given by
\begin{equation} \label{varphiRI}
\varphi_{\bf I} (u,v\,;\, q_{i})= \ell^{2}\frac{\bar{h}\Delta N_{\Delta}}{2^{2\Delta-2}} \int^{u}_{q_{i}} dq \Big[w^{1-\Delta}(1-w)^{2\Delta-1} + \Delta \frac{1+w}{1-w}B_{w}(1-\Delta,2\Delta)\Big]\,,
\end{equation}
where  we have restored $\ell$ and  $B_{w}$ denotes the incomplete Beta function 
\begin{equation} \label{}
B_{w}(a,b)\equiv \int^{w}_{0}dt~ t^{a-1}(1-t)^{b-1} = \frac{w^{a}}{a}F(a,1-b,;\, 1+a\,|\, w)\,, \nonumber 
\end{equation}
and  $N_{\Delta}$ was introduced before as  $N_{\Delta} = \bar{N}_{\Delta} \sin\pi \Delta= 2\big[\frac{2^{\Delta-2}\Gamma^{2}(\Delta)}{\pi\Gamma(2\Delta)}\big]^{2}\sin\pi \Delta$.  

When $q_{i} < v$, it is sufficient to take $q_{i}$  to be its minimum value $v$ and so the contribution is given by
\begin{align}    \label{}
\varphi_{\bf I}(u,v) & \equiv \varphi_{\bf I}(u,v\,;\, q_{i}=v) \\
&= {\textstyle 4\ell^{2}\Delta Q_{\Delta}\frac{B(2-\Delta,2-\Delta)}{1-\Delta}{\scriptstyle \tan(|u-v|)\sin^{2-2\Delta}(|u-v|) } F\big(1-\Delta,1-\Delta\,;\, \frac{5}{2}-\Delta\,|\, {\scriptstyle \sin^{2}(u-v) } \big) }\,. \nonumber 
\end{align}
\underline{Region {\bf III}}: Figure~\ref{pqrange} tells us that $q_{i} < v$ in this region. When $q_{i} > v- \frac{1}{2}$, the contribution to the dilaton deformation is given by 
\begin{align}    \label{}
\varphi_{\bf III} (u,v\,;\, q_{i}) &= -\ell^{2}\frac{\bar{h}\Delta^{2} }{2^{2\Delta-2}}N_{\Delta}  B(-\Delta,2\Delta)  \int^{v}_{q_{i}}  dq \frac{\cos(u+v-2q)}{\cos(u-v)} \nonumber \\
&= \ell^{2}\frac{\bar{h}\Delta^{2} }{2^{2\Delta-1}}N_{\Delta}B(-\Delta,2\Delta)\Big[ \tan (u-v)  - \frac{\sin(u+v-2q_{i})}{\cos(u-v)}\Big]\,. 
\end{align}
When $q_{i} <  v- \frac{1}{2}$, by taking $q_{i}$ to be its lowest value $v-\frac{\pi}{2}$,   the dilaton deformation becomes  
\begin{equation} \label{}
\varphi_{\bf III} (u,v) \equiv \varphi_{\bf III} (u,v\,;\, q_{i}=v-{\textstyle \frac{\pi}{2}})  = \ell^{2}\frac{\bar{h}\Delta^{2} }{2^{2\Delta-2}}N_{\Delta}B(-\Delta,2\Delta) \tan \mu\,. 
\end{equation}
\underline{Region ${\bf I}'$ and ${\bf III}'$}: 
Because of the phase assignments given in the appendix, one can also easily see that  
\begin{equation} \label{Primed}
\varphi_{{\bf I}'} (u,v\,;\, q_{i}) = -\varphi_{{\bf I}} (u,v\,;\, q_{i}+{\textstyle \frac{\pi}{2} } )\,, \qquad  \varphi_{{\bf III}'} (u,v\,;\, q_{i}) = -\varphi_{{\bf III}} (u,v\,;\, q_{i}+{\textstyle \frac{\pi}{2} }  )\,.
\end{equation}

As a result,
depending on the position of the initial time $\tau_{i} = 2q_{i} - \frac{\pi}{2}$ over the range $u-\pi <q_{i} < u$,   the contribution to the dilaton deformation may be summarized as follows:
\begin{equation} \label{Varphi} 
\varphi (u,v) = 
\left\{ \begin{array}{lcc}    
\varphi_{\bf I} (u,v\,;\, q_{i} )   & {\rm for}    &  v < q_{i}   < u   \\
\varphi_{\bf I} (u,v)  +\varphi_{\bf III} (u,v\,;\, q_{i} )  &    {\rm for}    &  u -\frac{\pi}{2}  < q_{i}   < v   \\
\varphi_{\bf I} (u,v) + \varphi_{\bf III} (u,v\,;\, q_{i}) +\varphi_{{\bf I}'}(u,v\,;\, q_{i}) & {\rm for}    &   v- \frac{\pi}{2}  < q_{i}   <  u -\frac{\pi}{2} \\
\varphi_{\bf I} (u,v) + \varphi_{\bf III} (u,v) +  \varphi_{{\bf I}'}(u,v) + \varphi_{{\bf III}'} (u,v\,;\, q_{i})& {\rm for}    &   u-\pi < q_{i}   < v-\frac{\pi}{2} 
\end{array}  \right.    \,.
\end{equation}
Over the range $q_{i} > u$, there is no contribution to $\varphi$ because of the causality, and over  the range $u-(n+1)\pi <q_{i} < u - n\pi $ for $n=1,2,3,\cdots$,  it  takes the same expression as the above, because of the periodicity of $\varphi$. In the next sections, we present various solutions by summing up  these contributions to $\varphi$ appropriately in the context of our physics.

\section{Transition from black holes to eternal traversable wormholes}\label{sec4}

In this section, we shall construct smooth solutions describing a transition from black holes  to eternal traversable wormholes. 
These are time-dependent  solutions and, in the next section,  we shall construct solutions describing  static eternal traversable wormholes.  

\begin{figure}[th]
\vskip-1cm
\begin{center}
\includegraphics[width=8.5cm]{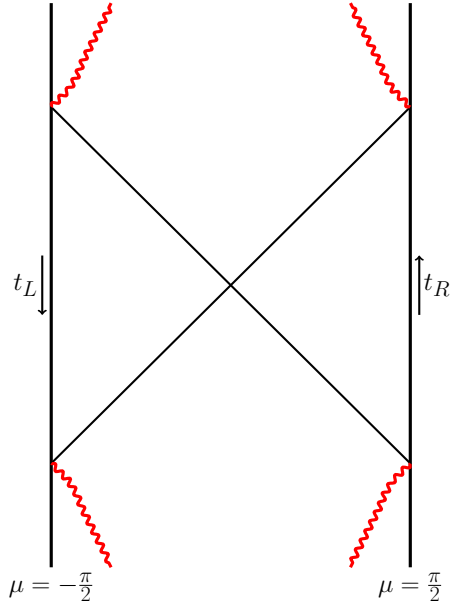}
\end{center}
\vskip-1cm
\caption{
\label{fig301}We draw the Penrose diagram for the AdS$_2$ black hole in $(\tau,\mu)$ space where the wiggly red lines represent the location 
of past and future singularities. 
We draw here the case where the singularities are timelike with $\phi_0/(L\bar\phi )=1/2$. 
}
\end{figure}

 We shall begin with a brief description of the black hole spacetime 
  \bea \label{DilatonSol0}
\phi= \phi_{\rm bh}(L,1,0) = \bar\phi \, L\,\,\frac{ \cos \tau }{ \cos \mu} \,,
\eea 
which will be our choice of initial  bulk configuration. In particular, this black hole spacetime is left-right symmetric under the exchange of $\mu \leftrightarrow -\mu$.
By
the coordinate transformation (\ref{coorb}) with $b=1$ and $\tau_B=0$,
%
%
one is led to the 
 black hole metric
%
 in (\ref{btz}) 
%
with the dilaton
%
$\phi= \bar\phi \, r$,
%
and  may identify  $L$ as the radius of black hole horizon. 
 Then, the Gibbons-Hawking temperature of the black hole is identified as 
\begin{equation} \label{}
T= \frac{1}{2\pi} \frac{L}{\ell^2}\,,
\end{equation}
and the entropy  and energy are given  respectively  by
\begin{align}    
S= S_0 +{\cal C} T\,,  \ \ \ \ E = \frac{1}{2} {\cal C}T^2\,,
\label{energybh} 
\end{align}
where ${\cal C} =\frac{\pi \bar\phi \ell^2}{2 G}$ and $S_0$ is the zero temperature contribution given by $S_0 = \phi_0/(4G)$.
The entropy can be written as a Bekenstein formula
\begin{equation} \label{}
S= \frac{\phi_0 +\bar\phi\, L} {4G}\,.
\end{equation}
The Penrose diagram of the black hole spacetime is depicted in Figure \ref{fig301}.  The singularities  are defined by $\Phi^2= \phi_0 + \phi=0$ where $\Phi^2$ corresponds to 
the radius squared of the transverse space assuming the viewpoint of the dimensional reduction from higher dimensions. When $\phi_0 > L \bar\phi$ ($\phi_0 < L \bar\phi$), the singularities
are timelike   (spacelike)  where $\phi_0$ is taken to be positive by definition. If $\phi_0 = L \bar\phi$, the singularities are lightlike.  In this note, we shall be interested in the 
case $\phi_0 \gg L \bar\phi$ with the timelike singularities. In this section, we shall consider $\tau > -\pi/2$ or $t  > -\infty$ (with $t$ denoting here the right Rindler time) such that one is  always away from the 
past singularities.

 In this two-sided black hole spacetime, the L-R boundaries are connected by a wormhole (so-called Einstein-Rosen bridge)  but  causally disconnected from each other. Thus  this  wormhole  is  not
 traversable. If one sends a signal from one boundary, it falls into the horizon hitting singularities. It never reaches the other side, so this wormhole is not traversable at all. The black hole is also eternal in the sense that it never evaporates spontaneously.
 
 Let us now turn to the boundary description introduced in the previous section. First, starting from our bulk description, we introduce the cut-off prescription
$ \phi|_{\partial M}= \frac{1}{\epsilon} \bar\phi \ell $
in both asymptotic regions of the L/R boundaries. Obviously this leads to the cut-off $r=\ell/\epsilon$ in the L/R asymptotic regions. Using the prescription 
$ds^2= -\frac{1}{\epsilon^2} d \tilde{u}^2$ at the cut-off,
one finds 
\be \label{tilut}
\tilde{u} =\mp \, t_{L/R} = \mp \, t 
\ee
where $-/ \tt +$ is respectively for the left/right boundary. Note that $t_R$ runs upward whereas $t_L$ runs downward, which is just our convention and in accordance with the time translation symmetries of our black hole geometry. Since $L/R$ boundaries are non-interacting with each other, there is such a freedom to choose these boundary times different. However in this note we shall use a single time $\tilde{u}$ as our boundary time in the end to make a coherent description including the case of L-R interactions. This time runs upward for both boundaries, which is required when the wormhole is embedded into a 
bigger system as a subsystem. Let us now check this solution with the Schwarzian  description in (\ref{sbdry}) where we set $g =0$  turning off the interaction term. 
The equations of motion in the Schwarzian theory can be solved 
\be
\tilde{u} =\mp \, t_{L/R}\,,
\ee
where we have used the relation 
\be
\tanh \frac{L t_{L/R}}{2 \ell^2}= \mp \tan \frac{\tau_{l/r}}{2}\,.
\ee
In this solution, we have used the SL(2,R) symmetries of  (\ref{sbdry}) to fix the overall scaling and a constant shift of time. Therefore one finds a perfect agreement of the bulk and the Schwarzian descriptions. 

\begin{figure}[th]
\vskip-1cm
\begin{center}
\includegraphics[width=8cm]{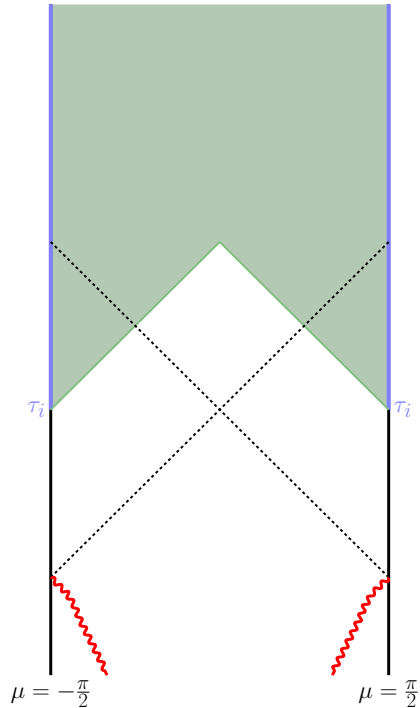} 
\end{center}
\vskip-1cm
\caption{
\label{fig401} We draw the Penrose diagram for the geometry describing the transition from  black hole to eternal traversable wormhole 
 in $(\tau,\mu)$ space.  
The double trace deformation will be turned on at $\tau=\tau_i$, which makes the wormhole traversable. 
}
\end{figure}

We now turn to the transitional solution from the black hole to an ETW spacetime. Here our construction  is based 
on the dilaton solution described in the previous section. We shall turn on L-R interaction of the previous section from $\tau= \tau_i$ in the background of black holes in which, later on, we shall
 set $\tau_i=0$. In Figure \ref{fig401}, we draw the region where the L-R interaction is effective in the bulk, which is shaded by green color in Figure \ref{fig401}.  The unshaded region, on the other hand, 
is not affected by the interaction as dictated by the causality of the spacetime. 

In the previous section, we construct the dilaton solution for the region $u > v $ or $\mu >0$. The solution in the other side $u < v$ or $\mu <0$ can be obtained using symmetry of the solution under the exchange $u \leftrightarrow v$ or $\mu \leftrightarrow -\mu$. This symmetry follows since our turned-on  source as well as the initial configuration are symmetric under the 
exchange of the left and the right. Here we assume that $\bar{h}$ remain constant once turned on, which will be justified later on.  Namely we assume
$\tau' \equiv \frac{d\tau}{d\tilde{u}}$ constant  when $q \ge q_i$.

Let us now first concentrate on the solution on the right side $u > v$ from regions {\bf III} and ${\bf III}'$ in Figure \ref{pqrange}, 
which will be denoted by $\varphi_h$.    As was described in the previous section, $\varphi_h$ is nonvanishing only when $q_i <  v$.   For  $ v  -\frac{\pi}{2} < q_{i}   < v $, one has
\begin{equation} \label{}
\varphi_h = \varphi_{\bf III} (u,v\,;\, q_{i} ) \,.
\end{equation}
%
%
For the next region of  $ v  -{\pi}< q_{i}   < v -\frac{\pi}{2} $, one has
\be
\varphi_h = \varphi_{\bf III} (u,v; v -\frac{\pi}{2} )  +\varphi_{\bf III'} (u,v\,;\, q_{i} ) = \varphi_{\bf III} (u,v\,;\, q_{i} ) \,.
\ee
One may continue this further.
Namely for $ v  -\frac{3\pi}{2} < q_{i}   < v-\pi $, one has
\be
\varphi_h = \varphi_{\bf III} (u,v; v -\frac{\pi}{2} )  +\varphi_{\bf III'} (u,v\,;\, v-\pi ) + \varphi_{\bf III} (u,v\,;\, q_{i}+\pi ) = \varphi_{\bf III} (u,v\,;\, q_{i} ) \,.
\ee
In this manner, one finds
\be
\varphi_h =\theta(v-q_i) \, \varphi_{\bf III} (u,v\,;\, q_{i} ) \,,
\ee 
which is valid only for the right side of $ u > v$. 
Using the left-right symmetry of the configuration, one has
\be
\varphi_h =\theta(u-q_i) \, \varphi_{\bf III} (v,u\,;\, q_{i} ) 
\ee 
for the left side of $u < v$. Note that this solution has the form of homogeneous solution in each side of the bulk and our construction of the solution is valid only
up to homogeneous contributions. To fix this homogeneous part properly, 
we use (\ref{eqphi}) and  obtain 
the corresponding energy momentum tensor $T^h_{ab}$ which is found to be nonvanishing and localized along the center $\mu=0$ with $\tau > \pi/2+\tau_i$. This implies the presence of an extra source term along the center. 
  This source term should be absent from the beginning
since we put the source term only at the L-R boundaries. Hence we need to remove this energy-momentum contribution  by placing an  opposite-signed source term 
along  the center $\mu=0$ with $\tau > \pi/2+\tau_i$. Thus one  has 
$\varphi_h =0$ in the end.

We next turn to the contributions from {\bf I} and $\rm \bf I'$ together with their related regions by $\pi$-periodicity.  For the right side of 
$u>v$, the dilaton is nonvanishing only when $q_i <u$.
Omitting the homogeneous 
contributions in (\ref{Varphi}) and using the relation in Eq.~(\ref{Primed}),  one has
\begin{equation} \label{Varphiff} 
\varphi_{\rm tw} (u,v;q_i) = 
\left\{ \begin{array}{lcc}    
\varphi_{\bf I} (u,v\,;\, q_{i} )   & {\rm for}    &  v < q_{i}   < u   \\
\varphi_{\bf I} (u,v)   &    {\rm for}    &  u -\frac{\pi}{2}  < q_{i}   < v   \\
\varphi_{\bf I} (u,v) -\varphi_{{\bf I}}(u,v\,;\, q_{i}+\frac{\pi}{2}) & {\rm for}    &   v- \frac{\pi}{2}  < q_{i}   <  u -\frac{\pi}{2} \\
0 & {\rm for}    &   u-\pi < q_{i}   < v-\frac{\pi}{2} 
\end{array}  \right.    \,,
\end{equation}
which is  for $ u-\pi < q_i  < u$.  When $q_i < u-\pi$, the corresponding solutions are determined by the $\pi$-periodicity
\bea
\varphi_{\rm tw} (u,v;q_i) =\varphi_{\rm tw} (u,v;q_i+n\pi) \,,
\eea 
where the integer $n$ is fixed by requiring $ u-\pi < q_i  +n \pi < u$. For the left side $u < v$, the dilaton contribution is nonvanishing only when  $ q_i < v$
 and determined by the relation $\varphi_{\rm tw} (u,v;q_i)= \varphi_{\rm tw} (v,u;q_i)$ where we use the left-right symmetry. One can check that there is no extra localized
  source term placed 
 along the center $u=v$ in this case.
 
 To complete the solution, we set $q_i =\pi/4$ or $\tau_i=0$ as announced already and choose $h$ such that\footnote{Of course, one can choose $\tau_i$ arbitrary turning on the double trace deformation at arbitrary $\tau_i$. With constant $h$, this then in general makes the boundary velocity $\tau'$ 
non constant in $\tilde{u}$ and the above construction is not applicable anymore. If $\tau_i$ is small enough, the bulk shows the phenomenon of  graviton oscillation  of eternal traversable wormhole, which will be described in section \ref{sec7}. }
 \be
 h=h_{\rm tw}(L)\equiv \bar\phi\frac{ 4\pi}{\Delta B(\Delta,\Delta)} \left(\frac{L}{\ell^2}\right)^{2(1-\Delta)}  \,,
 \label{htw}
 \ee
 where the scale $L$ is set by the initial black hole state. 
 Therefore our full solution describing the transition from the black hole to the  ETW spacetime 
 is given by
 \be
 \phi =\bar\phi \, L\,\,\frac{ \cos \tau }{ \cos \mu} +\varphi_{\rm tw}(u,v;\pi/4,L) \,,
 \label{bbhetw}
 \ee
 where we  used Eq.~(\ref{asympphi1}) for $\varphi_{tw}$ and introduced  a new notation 
 \be
 \varphi_{\rm tw}(u,v;\pi/4,L) \equiv \varphi_{\rm tw}(u,v;\pi/4)\big{|}_{h=h_{\rm tw}(L)}\,,
 \ee
 emphasizing the fact  that $h$ (or $g$) is tuned to a  particular strength  defined in (\ref{htw}), fixed by   the length scale $L$.
 In  asymptotic regions, the dilaton solution $\phi$ takes a simple form 
 \be
 \phi = \bar\phi  L \frac{\cos \tau}{\cos \mu} \theta(-\tau) +\bar\phi  L \frac{1}{\cos \mu} \theta(\tau)+ {\cal O}(\cos^{2\Delta}\mu ) \,,
 \ee
both for the left and the right sides.  This then leads to the boundary dynamics
\be
\phantom{aaaaa} \tilde{u}=  \theta(-\tau)  \frac{\ell^2}{L} {\rm arctanh} \, \sin \tau+ \theta(\tau) \frac{\ell^2}{L}\tau \ \ \ \  
  (\tau > -\pi/2)\,,
   \label{sbhetw}
\ee
 which is describing a transition from a thermal state to an ETW state. Of course this perfectly agrees with the result from  the 
 Schwarzian boundary description of  (\ref{sbdry}).

With this solution, one may send a signal from one to the other side without any obstruction.  The time delay sending the signal from one to the other side is given by\footnote{See \cite{Gao:2018yzk} where the delay of signals  for the  ETW state is investigated from  field-theoretic point of view.}
\be 
\Delta \tilde{u}=\frac{\ell^2 \pi}{L}+ \left( \frac{\ell^2}{L} {\rm arcsin }\, \tanh \frac{L \tilde{u}}{\ell^2}
-\tilde{u}
\right) \theta(-\tilde{u}) \,,
\label{delay}
\ee
where $\tilde{u}$ is the initial time for sending signals.
When $\tilde{u} >0$,  the reappearance time is simply given by $\pi \ell^2/L$, which is signaling that our boundary system is in the eternal traversable wormhole state. The future 
horizons of the bulk geometry disappear completely. In addition the future singularities defined by the relation $\phi_0+\phi=0$ disappear as well and the bulk solution becomes fully 
regular consequently.
We assume here $\phi_0 \gg \bar\phi L$,  which is required for the validity of our one-loop approximation. Therefore any information behind the would-be horizon becomes 
available
to the asymptotic regions of L-R boundaries. This is a rather nice example of the disappearance of horizons, which may be called as an evaporation of black hole 
to eternal traversable wormhole. Namely, the wormhole becomes fully transparent though there may be a significant  time delay for early enough signals.

Clearly this has a dual microscopic  CFT$_1$ description since  AdS/CFT correspondence is available for our 2d dilaton gravity \cite{Maldacena:2016upp}.
 In the boundary description, any early initial perturbation at $\tilde{u} <0 $ scrambles  as in  typical thermal systems  but, with the double-trace interactions turned on, the scrambled information reappears in a 
purified form with the time delay in (\ref{delay}). Of course, without the double-trace interaction turned on, this perturbation will scramble and effectively  disappear into the 
thermal bath. In the dual black hole spacetime, the perturbation falls into the horizon and eventually hit the future singularities. Such a fate of information will be saved by our
double-trace deformation as the  wormhole  becomes traversable and transparent. 

\section{Eternal traversable wormholes}\label{sec5}
In this section  we  consider eternal traversable wormholes. Later we shall also construct the transitional solution from an eternal traversable wormhole to a black hole.  As was shown in Ref.~\cite{Maldacena:2018lmt}, 
%
the solutions in the Schwarzian theory for these wormholes are given by
\begin{equation} \label{}
\frac{d}{d\tilde{u}}\tau (\tilde{u}) = \tau'(\tilde{u}) =  const.
\end{equation}
where $\tau$ is the gauge fixed global time field in the Schwarzian theory as $\tau(\tilde{u})= \tau_{l}(\tilde{u})=\tau_{r}(\tilde{u})$.   
Without turning on the double trace deformation,   there is no wormhole solution and so the above constant solution should vanish when the coupling $g$ goes to zero.  To see this, let us recall that  the equation of motions in the boundary Schwarzian theory  in the chosen gauge becomes
\begin{equation} \label{eqsch}
\Big[\bar{\phi}\tau'+ \bar\phi \frac{1}{\tau'}\Big(\frac{\tau''}{\tau'}\Big)' - \frac{g\Delta}{2^{2\Delta}} \tau'^{2\Delta-1} \Big]' =0\,.
\end{equation}
Among the solutions of  the above equation,  those  with  $SL(2,{\bf R})$ charge $Q_{0}$ fixing  condition  are determined by
\begin{equation} \label{taup1}
\bar{\phi}\tau' - \frac{g\Delta}{2^{2\Delta}} \tau'^{2\Delta-1}=0\,.
\end{equation}
Then, one can see that the solution is given by
\begin{equation} \label{taup2}
\tau'(\tilde{u}) = \Big(\frac{ g\Delta}{2^{2\Delta}\bar\phi}\Big)^{\frac{1}{2(1-\Delta)}} \equiv  \frac{L_g}{\ell^2} \,. 
\end{equation}
%
We would like to understand  this solution  from the bulk Einstein-dilaton theory.  To compare the bulk side with the boundary Schwarzian one, let us start the background configuration of the dilaton field as $\phi_{bg} =0 $ instead of black holes in Eq.~(\ref{DilatonSol0}).  Now, one may ask the effect of the double trace coupling in this setup, which is turned on indefinitely long ago.  Our goal is to compute the dilaton field deformation  $\varphi$ for such a double trace interaction. 
%
%
%

Since the dilaton deformation in each region is computed in the previous section for a given $\tau_{i}$, it would be sufficient to organize the deformation expression appropriately to identify eternal traversable wormholes in the bulk view point. As was discussed in the previous section,  the initial time $\tau_{i}=2q_{i}-\frac{\pi}{2}$, turning on  the double trace interaction,   is introduced  for the given  background to deform black holes to wormholes. There would be no such a special initial time $q_{i}$ in the case of eternal wormholes. This means that the initial time effect should be removed in some way. We show that this  can be achieved by averaging the expression of the dilaton deformation in Eq.~(\ref{Varphi}).  By averaging, we means that the turning-on time $q_{i}$  is integrated  and divided by its (quasi-)periodicity range  in the following sense
\begin{equation} \label{}
\varphi_{avg}(u,v) = \frac{1}{\pi} \int^{u}_{u-\pi}dq_{i} ~ \varphi(u,v)\,.
\end{equation}
By using the results in Eq.~(\ref{Varphi}), one can see that  the contribution from the regions {\bf I} and ${\bf I}'$ is given by
\begin{align}    \label{}
\varphi_{{\bf I},\, avg}(u,v) &=  \frac{1}{\pi}\bigg[  \int^{u}_{v}dq_{i}~ \varphi_{\bf I} (u,v\,;\, q_{i}) +  \int^{v}_{u-\frac{\pi}{2} }dq_{i}~ \varphi_{\bf I} (u,v) \nonumber \\
& \quad  + \int^{u-\frac{\pi}{2}}_{v-\frac{\pi}{2} }dq_{i}~ \Big(\varphi_{\bf I} (u,v)   - \varphi_{\bf I} (u,v\,;\, q_{i}+{\textstyle \frac{\pi}{2}} ) \Big)
 + \int^{u-\frac{\pi}{2}}_{v-\frac{\pi}{2} }dq_{i}~ \Big(\varphi_{\bf I} (u,v)   - \varphi_{\bf I} (u,v) \Big)  \bigg] \nonumber \\
&=  \frac{1}{2}\varphi_{\bf I}(u,v)\,, \nonumber 
\end{align}
where we have used the result  in Eq.~(\ref{Primed}) that  the value of dilaton deformation on the   region  ${\bf I}'$ is given by the negative of its corresponding ones on the  region {\bf I}.   

As was explained in the previous section, it turns out that the contribution from $\varphi_{\bf III}$  and $ \varphi_{\bf III}'$ leads to the extra contribution to stress-tensor given in the form of  the homogeneous solution, and so it should be removed in our final result. 
Therefore, the final result from averaging is given simply by 
\begin{equation} \label{}
\phi_{\rm etw} (u,v; L_g)= \frac{1}{2}\varphi_{\bf I}(u,v)\big{|}_{h= h_{\rm tw}(L_g)}\,.
\end{equation}
This is our bulk solution corresponding to the Schwarzian one given in (\ref{taup2}).

Now, let us check whether our bulk solution  is matched to the Schwarzian one. From the cut-off prescription in Eq.~(\ref{bcondition}), one can see that
\begin{equation} \label{}
\frac{d\tau}{d\tilde{u}}=  \frac{1}{\ell}\frac{\sin \xi}{\epsilon} \,, \qquad \phi_{\rm etw}\Big{|}_{\partial M}  
= \frac{\bar{\ell\,\phi}}{\epsilon} 
= 
 \frac{1}{2}\varphi_{\bf I}\Big|_{\partial M}\,,
\end{equation}
where $\xi\equiv \frac{\pi}{2}-\mu$.
Using the asymptotic expression of the dilaton deformation in Eq.~(\ref{asymp}) and restoring $\ell^{2}$ in $\varphi_{\bf I}$,   one obtains 
\begin{equation} \label{}
\tau'=\frac{d\tau}{d\tilde{u}}= \frac{1}{2}\frac{\bar{h}\Delta^{2} N_{\Delta}}{2^{2\Delta-2}\bar{\phi}} B(1-\Delta,2\Delta)= \frac{h \tau'^{2\Delta-1}}{4\pi\bar{\phi}}\frac{\Delta \Gamma^{2}(\Delta)}{\Gamma(2\Delta)} \,,
\end{equation}
where we used the relation  in Eq.~(\ref{barh}). 
By using the relation between the coupling $g$ and $h$ in Eq.~(\ref{gtoh}), one can rewrite the above as
\begin{equation} \label{}
\tau' -  \frac{g \Delta}{2^{2\Delta}\bar\phi} \tau'^{2\Delta-1} =0\,,
\end{equation}
which shows us  the complete matching of the bulk expression to  the boundary Schwarzian one in 
(\ref{taup1}) and~(\ref{taup2}).

Let us now construct solutions describing  transition from  eternal traversable wormholes to black holes. We begin with an eternal traversable wormhole 
and turn off the double trace interaction at $\tilde{u}=0$. Repeating our construction in a similar manner, one finds
\be
\phi = \phi_{\rm etw}(u,v;L_g) -\varphi_{\rm tw}(u,v;\pi/4,L_g)
\label{betwbh}
\ee   
 In  asymptotic regions, the dilaton solution $\phi$ takes a simple form 
 \be
 \phi = \bar\phi  L_g \frac{\cos \tau}{\cos \mu} \theta( \tau) +\bar\phi  L_g \frac{1}{\cos \mu} \theta(-\tau)+ {\cal O}(\cos^{2\Delta}\mu ) \,,
 \ee
and one can see the development of horizon and future singularities. 
This then leads to the boundary dynamics
\be
\phantom{aaaaa} \tilde{u}=  \theta(\tau)  \frac{\ell^2}{L_g} {\rm arctanh} \, \sin \tau+ \theta(-\tau) \frac{\ell^2}{L_g}\tau \ \ \ \  
  (\tau < \pi/2)\,,
  \label{setwbh}
\ee
 which is describing a transition from  an ETW state to a thermal state \cite{Maldacena:2018lmt}. Of course this agrees with the result from  the 
 Schwarzian boundary description of  (\ref{sbdry}). In this Schwarzian description, one find  that the boundary solution (\ref{setwbh}) is simply given 
 by the time reversal transformation of our previous solution (\ref{sbhetw}) (with $L=L_g$) describing the transition from a thermal to an ETW state. On the other hand, the corresponding bulk solutions
 in (\ref{betwbh}) and (\ref{bbhetw}) do not show any time reversal symmetry since bulk propagation of (on and off) interactions should be dictated by causality. 

\section{Matter excitations}\label{sec6}
In this section we would like to discuss bulk solutions describing matter excitations above the ETW state denoted as 
$|{\rm ETW}\rangle$.  Here we consider a 
scalar field $\tilde\chi$ that is dual to a scalar  operator with dimension $\tilde\Delta$ and
construct its full back-reacted solutions.
The scalar field equation can be solved by \cite{Spradlin:1999bn}
\bea
\tilde\chi= \sum^\infty_{n=0}\bigl[  \, \lambda_n \, \chi_n(\mu,\tau,\tilde{\Delta})+\lambda^*_n \, \chi^*_n(\mu,\tau,\tilde{\Delta})\,  \bigr]\,,
\label{bulkscalar}
\eea
where 
we 
used the mode functions in (\ref{chimode}).
In operator forms, one may write it as
\be
\hat{\tilde\chi}= \sum^\infty_{n=0}\bigl[ \,  a_n \, \chi_n(\mu,\tau,\tilde{\Delta})+ a^\dagger_n \, \chi^*_n(\mu,\tau,\tilde{\Delta}) \,\bigr]\,,
\ee 
where $a_n$ is the lowering operator annihilating the ETW state satisfying commutation relation $[a_m, a^\dagger_n]=\delta_{mn}$.  
The corresponding  excited state can be given by
a coherent state
\be
|\vec{\lambda} \rangle = e^{-\frac{{\vec{\lambda}}^2}{2}} e^{\sum^\infty_{n=0}\lambda_n a^\dagger_n}\, |{\rm ETW}\rangle\,,
\ee
satisfying $a_n |\vec{\lambda} \rangle =\lambda_n |\vec{\lambda} \rangle $ with $\langle \vec{\lambda}|\vec{\lambda} \rangle=1$.  Note that
the bulk scalar field $\tilde{\chi}$ in (\ref{bulkscalar}) follows from   an expectation value of the bulk operator,
$\langle \vec{\lambda}| \hat{\tilde\chi}|\vec{\lambda} \rangle$, which may be regarded as a dictionary of the 
 AdS/CFT correspondence.

To see  what this deformation describes, we need to look at the dilaton part whose identification will 
complete the fully back-reacted gravity solution.  Here we shall consider only $n=0$ case choosing $\lambda_n =\delta_{n0}$ for the 
sake of  illustration.
One may write this $n=0$ scalar solution as
\bea
\tilde\chi =\gamma  \cos^\Delta \mu \cos \tilde\Delta 
\tau  \,,
\eea
with 
\be
\gamma=  {\textstyle \frac{ 2^{\Delta}\Gamma(\Delta)}{\sqrt{\pi\Gamma(2\Delta)}}}  \,.
\ee
 The corresponding dilaton solution can be found as \cite{Bak:2018txn}
\begin{align}    \label{}
\varphi_{\rm matt} =  
 -\frac{1}{4}\frac{
 \gamma^2 \tDelta }{1+2\tDelta} \cos^{2\tDelta} \mu \cos 2\tDelta  \tau 
-
\frac{1}{4}\gamma^2 \tDelta \cos^{2\tDelta} \mu \,\, F \Big(\tDelta, 1\,;\,  \frac{1}{2} \,\Big|\, \sin^2 \mu\Big)\,,
\end{align}
where  
$F(a,b\,;\,c\,|\,z)$ denotes the hypergeometric function \cite{grad}. 
Together with the ETW part, described in  the previous section,
 the full dilaton solution reads
\be
\phi = \phi_{\rm etw} + \varphi_{\rm matt}  \,.
\label{mattde}
\ee 
To see its asymptotic structure in the region $\mu\rightarrow \pm\frac{\pi}{2}$,  we shall use the following relation
\begin{equation} \label{hyperg}
F\Big(\tDelta, 1\,;\,  \frac{1}{2}\,\Big|\, \sin^2 \mu\Big) = \frac{1}{1+2\tDelta}\,\, F\Big(\tDelta, 1\,;\,  \tDelta +\frac{3}{2}\,\Big|\, \cos^2 \mu\Big)
 +\frac{\Gamma(\frac{1}{2})\Gamma(\tDelta +\frac{1}{2})}{\Gamma(\tDelta) \cos^{2\tDelta} \mu }|\tan \mu|\,.
\end{equation}
In the asymptotic region,  the solution  becomes 
\bea
\phi =  \frac{g\Delta}{2^{2\Delta}} \tau'^{2\Delta-1} \frac{1}{ \cos \mu} 
 - 
\frac{\tDelta}{2}
\frac{1 }{\cos \mu}
  + {\cal O}(\cos^{2\Delta}\mu)+{\cal O}(\cos^{2\tDelta}\mu)\,.
\eea
This leads to the condition
\be 
\bar{\phi}\tau' =  \frac{g\Delta}{2^{2\Delta}} \tau'^{2\Delta-1}  -\frac{\tDelta}{2} \,,
\ee
which solves the equation of motion in (\ref{eqsch}) and   is consistent with the discussion in  \cite{Maldacena:2018lmt}. The effective boundary velocity $\tau'$ 
will change slightly
due to the matter excitations.
Thus we find here the boundary Hamiltonian remains by our matter excitation in the above.
Only the state is deformed by these matter excitations.  If $\phi_{\rm etw}$ is replaced by the black hole $\phi_{\rm bh}$ in (\ref{mattde}), the corresponding solutions describe
excitations above the thermal vacuum where any excitations will decay away exponentially describing thermalization  \cite{Bak:2018txn, Bak:2017xla, Bak:2017dkj}. 
This thermalization also represents general scrambling of thermal system where information is dissipated away into thermal bath. Thus basic properties our ETW state are fundamentally different 
from those of  the black hole state.

\section{Graviton oscillations}\label{sec7}
In this section, we consider bulk solutions describing bulk graviton oscillations above the ETW state. The relevant system is still described by the boundary effective action in 
(\ref{sbdry}) including the L-R interactions. With its equation of motion in (\ref{eqsch}), we consider a small oscillation around the ETW solution given by
\be
\tau' = \tau'_0 + \delta \tau'  \,,
\ee
where $\tau'_0 =L_g /\ell^2$ denotes the solution in Eq.~(\ref{taup2}). The equation of motion is then reduced to 
\be
(\delta \tau')''  + \omega^2_g \, \delta \tau'  =0 \,,
\ee
to the leading order with an oacillation frequency $\omega_g = \sqrt{2(1-\Delta)} \, \tau'_0 =\bar{\omega}_g \, \tau'_0$ \cite{Maldacena:2018lmt}. Its solution is given by
\be
\delta \tau' = A \sin \omega_g \tilde{u}  =  A \sin {\bar\omega}_g \tau +\cdots \,,
\label{graviton}
\ee
where $\cdots$ are denoting higher order terms.

Here we would like to find the corresponding bulk solution following construction of sections \ref{sec3}, \ref{sec4} and \ref{sec5}. We use general 
\be
\bar{h}(\tau)= h (\tau')^{2\Delta -1} \,,
\ee
which is time-dependent. To the leading order of perturbation, this $\bar{h}$ is expanded as
\be
\bar{h} =h  (\tau_0')^{2\Delta -1} \left( 1+ (2\Delta -1) \frac{\delta \tau'}{\tau'_0}\right) \,.
\ee 
To obtain the corresponding bulk dilaton solution, we shall use the above expression 
(\ref{graviton}) and replace $q$-dependent $\bar{h}$ inside the integrand of the $q$-integration in (\ref{Dilaton}). We then would like to  see if the resulting bulk solution is consistent with the above boundary description.  The result in the asymptotic regions
\be
\phi=\varphi_{\rm tw}(u,v;q_i) + \delta \varphi_g(u,v;q_i) \,, 
\ee 
where
\be
\delta \varphi_g= \frac{\bar\phi \ell^2}{\cos \mu}\Big[ A \sin \bar{\omega}_g \tau +(\bar{\omega}_g\tth-\tth 1) \sin\big( (\bar{\omega}_g\tth +\tth1) \tau_i\tth-\tth\tau \big)
\tth-\tth (\bar{\omega}_g\tth+\tth1) \sin\big( (\bar{\omega}_g\tth-\tth 1) \tau_i \tth+\tth \tau \big)
\Big]+\cdots \,.
\ee
Then by averaging over $\tau_i$ in an appropriate manner, one is led to
\be
\phi=\phi_{\rm etw}(u,v;L_g) + \delta \varphi_g(u,v) = \frac{\bar\phi \ell^2}{\cos \mu} \left( 
\tau'_0 + A \sin \bar{\omega}_g \tau
\right) +\cdots 
\ee
in the asymptotic regions. This then leads to the boundary solution in (\ref{graviton}). This bulk solution describes a pure gravitational oscillation above the ETW state, which is not 
directly related to the matter excitations 
of the previous section. 
This, together with the result in the previous section,  shows that the ETW state is gapped indeed.

\section{Probing eternal traversable wormholes}\label{sec8}
\begin{figure}[th]
\vskip-0.1cm
\begin{center}
\includegraphics[width=11cm]{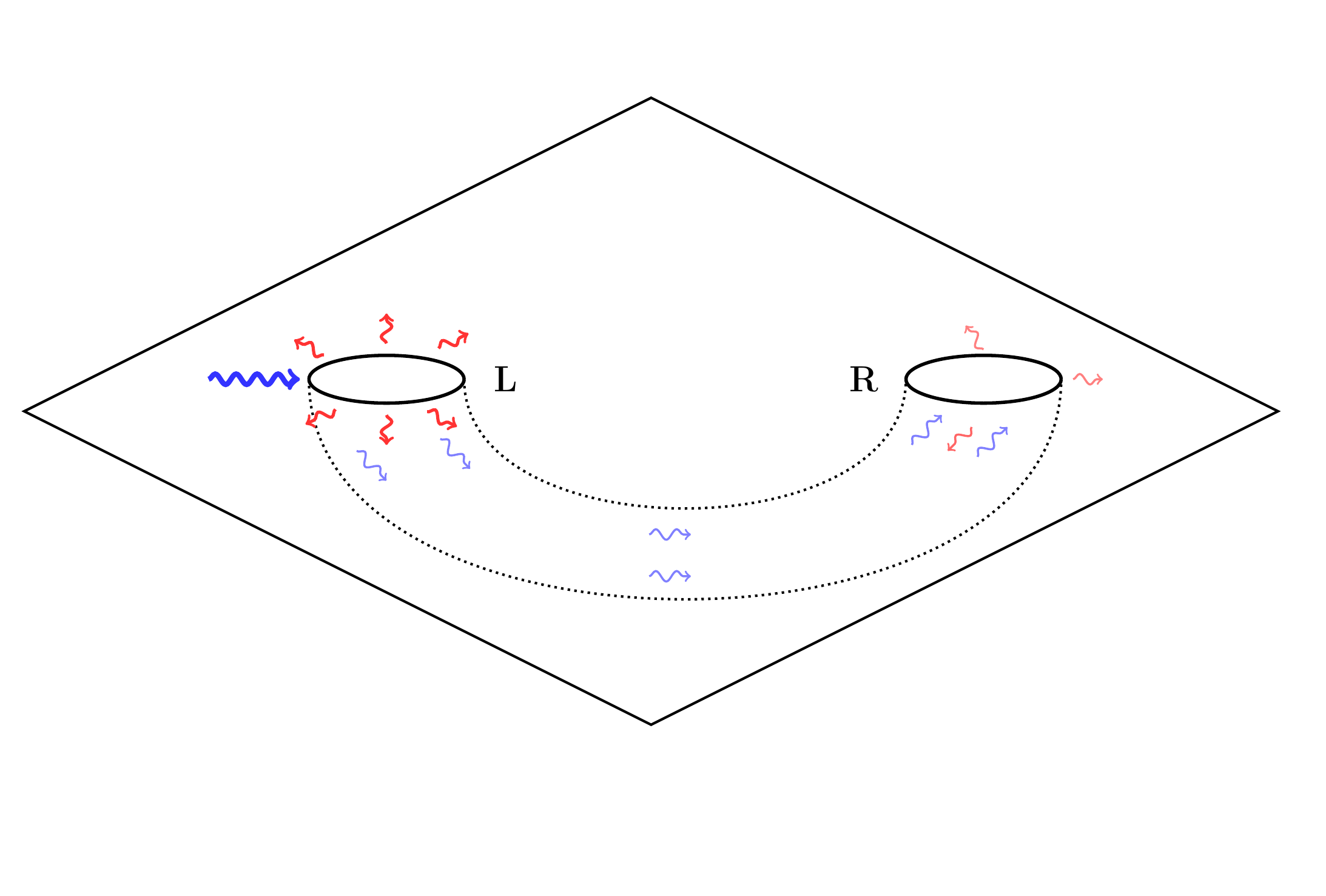} 
\end{center}
\vskip-1.0cm
\caption{
\label{fig501} \small We draw 4d eternal traversable wormhole connecting two magnetically charged holes. In the scattering problem, the incident wave is absorbed into 
one end of the wormhole, traveling to the other side. In the other side, there may be reflected wave as well as  transmitted wave to the asymptotic region in general.  In this sense,
the relevant scattering problem is unconventional.  
}
\end{figure}

In this section, we analyze various properties of the ETW state. We shall first describe how  information can be transferred  from one side to the other.
We then briefly  discuss 4d scattering problem through an eternal traversable wormhole where our 2d wormhole times a two sphere forms a 4d 
wormhole geometry as depicted in Figure \ref{fig501}.

Basically the ETW state is gapped as was shown in the previous sections. This gap makes the system non chaotic. Hence any perturbation  becomes 
non scrambling and will not be dissipated away, which makes the boundary system fundamentally different from the thermal system.

In thermal systems in general, any perturbation  will be dissipated away and the corresponding information will be lost to the thermal bath eventually,
which completely blocks any transfer of information from one  to the other side.  Indeed this fact can be checked with our gravity description where the left and right sides 
are causally disconnected from each other. The basic properties of the ETW system are precisely opposite to those of black holes. The gap makes the wormhole transparent
and traversable. There is no horizon as was discussed previously. In sending a signal from one side to the other, it takes a time $\Delta \tilde{u}= \frac{\pi \ell^2}{L_g}$. 
This may be explicitly realized as a bulk solution turning on the time dependent Janus perturbation \cite{Bak:2018txn, Bak:2003jk, Bak:2007jm}. A perturbation in one side
reappears in the other boundary, which verifies the reappearance time in the above. 

Let us now briefly discuss  the 4d traversable wormhole in  \cite{Maldacena:2018gjk} where two magnetically charged holes are connected by traversable 
wormholes. There in near extremal regions, the geometry can be approximated by AdS$_2\times S^2$ where AdS$_2$ part may be connected to the eternal traversable wormholes
in section \ref{sec5}. In this setup, one has 4d asymptotic region and may ask how eternal traversable wormholes look like if it is probed by a 4d massless scalar field
for instance. In particular one may ask an absorption cross section of one end of an  traversable wormhole to see how much waves falls through the wormhole out of the 
incident fluxes. This situation is depicted in Figure \ref{fig501}. In case of a black hole, the low energy scattering is well known and the absorption cross section is given by
the horizon area, $\sigma_{abs}= 4\pi r^2_+$ where $r_+$ is denoting the horizon radius defined by the largest zero of $H(r_+)=0$ in the 4d metric
\be
ds_{4d}^2 = - H(r_{4d}) dt_{4d}^2 +\frac{dr_{4d}^2}{H(r_{4d}) } +r_{4d}^2 ds^2_{S^2} \,, 
\ee
where $ds^2_{S^2}$ is the metric element  for a unit two-sphere.
In case of long traversable wormholes, the outer region of a hole in one side is well approximated by the above metric when the two holes are well separated. We expect the scattering cross section in the low energy limit is given by  $\sigma_{abs}\sim  4\pi r^2_+$ where now $r_+$ is no longer the horizon radius since there is no horizon in case of the eternal traversable wormhole.  Of course the absorbed flux will reappear to the other side since the wormhole is transparent. Then there will be a scattering problem from the wormhole to the outside region as well where we expect some part will be reflected back to the wormhole region.  Further study is required in this direction.

\section{Discussion}
In this paper, we have constructed several bulk solutions regarding eternal
traversable wormholes 
in 2d dilaton gravity with matter. The
basic strategy is to compute the deformation of the dilaton due to
double trace interaction between two AdS$_2$ boundaries by analytically
solving the equation of motion to the leading order in the interaction
parameter.
By turning on the double trace interaction at a particular time and keeping
the interaction indefinitely, we were able to obtain a time-dependent solution
which describes the transition of an initial AdS$_2$ black hole to an eternal
traversable wormhole
which is free of any singularity. It was also possible to construct
time-independent static ETW solutions by averaging the turning-on time 
of the interaction. On top of the ETW state, we have
considered matter excitations as well as bulk graviton oscillations,
which show that the ETW state is gapped and non-chaotic; any
perturbation becomes non-scrambling and will not be dissipated away, 
which makes the boundary system fundamentally different from the thermal 
system. We have explicitly checked that all these bulk constructions 
completely match with the boundary ones obtained in the Schwarzian 
boundary theory.

In the time-dependent solution where the initial black hole is 
transparentized into eternal
traversable wormhole, there are only past singularities, while 
future singularities no longer exist. Horizons disappear thanks to the 
double trace interaction between two boundaries, making the wormhole fully
transparent. In order to see more explicitly how the information behind the
would-be horizon comes out to the asymptotic regions, suppose that one 
sends a signal from one boundary to the other at a boundary time $\tilde u<0$
which is earlier than the time that the boundary interaction is turned on.
Then, the signal is initially sent in the black hole configuration. 
Note that $|\tilde u|$ can be very large. In this case, 
information would get dissipated away almost completely into the thermal bath.
Nevertheless, once the interaction is turned on, future singularities
disappear. There is no longer any obstruction that prevents signals
behind the would-be horizon from reaching the asymptotic region of
the other side, rendering the wormhole transparent. 

It would be illuminating to see more explicitly how long it 
takes for the signals sent before transparentization to reach the other side. 
The traveling
time of the signal is given by the time delay $\Delta \tilde{u}$ in 
\eqref{delay}. 
It is longer than that of the pure eternal traversable wormhole by the second term in \eqref{delay}
which is positive and monotonically increases as $\tilde u$ decreases. 
This extra time delay manifests that the signal has been sent in the black
hole configuration. 
Let us define $\tilde u_f$ as the boundary time that the signal sent at 
$\tilde u\,(<0)$ reaches the other side., i.e., 
$\tilde u_f = \tilde u + \Delta\tilde u $. 
If signals are sent periodically from $\tilde u = -\infty$ 
to $\tilde u =0$ with a constant time interval,
then all the signals would reappear at the other side during a finite period 
of time $\frac{\ell^2 \pi}{2L} < \tilde u_f < \frac{\ell^2 \pi}{L}$. 
In particular, when the signals start to come out at 
$\tilde u_f = \frac{\ell^2 \pi}{2L}$, 
one would see an initial sharp peak of signals since
$\frac{d \tilde u_f}{d \tilde u}$ vanishes  as $\tilde u \rightarrow -\infty$.
In this way, all the information may be recovered
after evaporation of the black hole to eternal traversable wormhole. The solutions
considered in this paper might provide an explicit
example how the information paradox could be evaded. Further study is needed
in this direction.


\section*{Acknowledgement}
D.B. was
supported in part by
NRF Grant 2017R1A2B4003095, and by Basic Science Research Program
through National Research Foundation funded by the Ministry of Education
(2018R1A6A1A06024977).
S.-H.Y. was supported by the National Research Foundation of Korea(NRF) grant with the grant number NRF-2018R1D1A1A09082212.


 
\appendix


\section{Phase assignments of $K_{L/R}$ } \label{AppA}
\renewcommand{\theequation}{A.\arabic{equation}}
  \setcounter{equation}{0}
 
 In this appendix,  we set $\ell =1$ and revisit the two-point (Hadamard) function in the covering space of global AdS$_{2}$ denoted usually as CAdS$_{2}$, in order to argue that we need a certain phase assignment in the two point function  in CAdS$_{2}$.  In the main text,   we have adopted  two kinds of bulk to boundary functions $K_{L/R}$ as  
\begin{equation} \label{}
K_{L/R}(\tau- \tau',\mu) = \lim_{\mu'=\mp \frac{\pi}{2}}\frac{1}{(\cos\mu' )^{\Delta}}  \Big\langle  \chi(x) \chi(x')  \Big\rangle\,,   
\end{equation}
where the upper sign is taken for $K_{L}$ and the lower one  for $L_{R}$.  Generically, one may set phases of  each $K_{L/R}$ as 
\begin{equation} \label{}
K_{L/R}(\tau \pm i\epsilon)  = e^{\pm i\nu_{L/R}} |K_{L/R}(\tau \pm i\epsilon) |\,,
\end{equation}
where the upper and lower signs of $\nu_{L/R}$ are taken from the usual $i\epsilon$-prescription convention~\cite{Gao:2016bin,Bak:2018txn}.  These phase assignments lead to the  expression of $K^{+}_{L/R}$ in Eq.~(\ref{phaseF}), while the additional $\theta(-d_{L/R})$ functions are inserted in $K^{-}_{L/R}$ in order to ensure the causality in the retarded functions. In the Rindler wedge case~\cite{Gao:2016bin,Bak:2018txn}, these $\theta$-function was implemented by the $i\epsilon$-prescription, but  in our phase assignments   it needs to be inserted as in Eq.~(\ref{phaseF})  since we may have non-vanishing unphysical result  of $K_{L/R}$  for the space-like separation.

To argue our choice of  the appropriate phase, $\nu_{L/R}$, let us remind the bulk two-point function.  As was shown in Ref.~\cite{Spradlin:1999bn}, the positive frequency mode solutions of a massive scalar  field $\chi$ with its mass $m^{2} =\Delta(\Delta -1) $ are given by
\begin{equation} \label{chimode}
\chi _{n}(\mu,\tau,\Delta) = {\textstyle  2^{\Delta-1}\Gamma(\Delta)\sqrt{\frac{n!}{\pi\Gamma(n+2\Delta)}} } ~ \cos^{\Delta}\mu~   C^{\Delta}_{n}(\sin \mu)~ e^{-i(n+\Delta)\tau} \,,   \quad n=0,1,2,\cdots \,, 
\end{equation}%
where $C^{\Delta}_{n}$ is the Gegenbauer function~\cite{grad}.  And then, the two-point function $\frac{1}{2} \langle  \{\chi(x),\chi(x')\}  \rangle   = {\rm Re}\sum^{\infty}_{n=0}\chi_{n}(x) \chi^{*}_{n}(x')$ could read from
\begin{equation} \label{Summ}
 \sum^{\infty}_{n=0}\chi_{n}(x) \chi^{*}_{n}(x')=   \frac{\Gamma^{2}(\Delta)}{4\pi \Gamma(2\Delta)}\Big[\Big(\frac{2}{\sigma}\Big)^{\Delta}F\Big(\Delta,\Delta\,;\, 2\Delta\,\Big|\, - \frac{2}{\sigma}\Big) \Big]\,,
\end{equation}
where the $SL(2,{\bf R})$ invariant distance $\sigma$ is defined by $\sigma\equiv \frac{\cos(\tau-\tau'-i\epsilon) - \cos(\mu-\mu')}{\cos\mu\cos\mu'}$. To arrive at the above final expression,  the following summation formula of the Gegenbauer function~\cite{prud}  may be used 
\begin{align}    \label{}
&\quad \sum_{n=0}^{\infty}\frac{n! \Gamma(2\Delta)}{\Gamma(n+2\Delta)}~ t^{n} C^{\Delta}_{n}(\cos x)C^{\Delta}_{n}(\cos y)   \nonumber \\
& = {\textstyle  \Big[\frac{1}{1-2t\cos(x-y) +t^{2}}\Big]^{\Delta}F(\Delta,\Delta\,;\, 2\Delta\,|\, - \frac{4t\sin x\sin y}{1-2t\cos(x-y)+t^{2}}) }\,,
\qquad |t| < 1\,,
\end{align}
with the overall multiplication of the factor  $t^{\Delta}$ in both sides of the equality,  and then one may take $x=\frac{\pi}{2}-\mu$, $y=\frac{\pi}{2}-\mu'$ and $t=e^{-i(\tau-\tau'-i\epsilon)}$. Here, $\epsilon >0 $ is inserted to satisfy the condition $|t| <1$ or to ensure the convergence of the infinite summation.

As is clear from the expression of $\chi_{n}$ in~(\ref{chimode}) with $\Delta \notin {\bf Z}$, the product of $\chi$ and $\chi^{*}$ satisfies 
\begin{equation} \label{}
\chi_{n}(x)\chi^{*}_{n}(x') \propto   e^{-i(n+\Delta)(\tau-\tau')}\,.
\end{equation}
This shows us that the left hand side of the equality (LHS) in~(\ref{Summ}) cannot be periodic in $\tau-\tau'$, while the right hand side of the equality (RHS) is periodic in $\tau-\tau'$ with the periodicity $2\pi$. In fact, the left hand side is quasi-periodic with the phase factor $e^{-i2\pi \Delta }$ for the $2\pi$-shift of $\tau-\tau'$. This mismatch may be traced back to the ambiguity in the choice of the phase in the expression of  $[1-2t\cos(x-y) +t^{2}]^{-\Delta}$, whenever it is evaluated beyond a single period of $\tau-\tau'$. To resolve this mismatch, we introduce appropriate phases in the  final expression of the two-point functions (RHS)  in such a way that it is consistent with the above quasi-periodicity of $\chi_{n}(x)\chi^{*}_{n}(x')$ (LHS). 
In other words, we add appropriate phases to the final expression of two-point functions to exhibit correctly their  quasi-periodic property.  With the consideration of the  boundary time directions, the quasi-periodicity may be implemented as 
\begin{equation} \label{}
K_{L}(\tau +2\pi)  = e^{i2\pi\Delta } K_{L}(\tau)\,, \qquad  K_{R}(\tau +2\pi)  = e^{-i2\pi\Delta } K_{R}(\tau) \,.
\end{equation}

There is another contribution to the phase $\nu_{L/R}$, whenever  $d_{L/R}$  in Eq.~(\ref{dfunc}) is negative. On the range of one periodicity $ -\pi < \tau < \pi$,  the phase factor $e^{ \pm i\pi \Delta }$  appears,  whenever $d_{L/R} < 0$, as 
\begin{equation} \label{}
K_{L/R} (\tau \pm i\epsilon) = e^{\pm i\pi \Delta} |K_{L/R}( \tau \pm i\epsilon)|\,.
\end{equation}
Our initial phase assignment  for  the range $ v<q<p$   ($d_{L} >0$ and  $d_{R} <0 $) may be taken as 
\begin{equation} \label{}
\nu_{L} =0\,, \qquad \nu_{R} = \pi \Delta\,.
\end{equation}
According to the tables~\ref{Tab1} and~\ref{Tab2} together with the above quasi-periodicity of $K_{L/R}$, it seems natural, then, to choose the phase assignment of  $K_{L/R}$  functions,  as is given  in  the following tables~\ref{Tab3} and~\ref{Tab4}.
\setcounter{table}{0}
\renewcommand{\thetable}{\color{blue}{\bf C}}
\begin{table} [htb]
\begin{center}
\begin{tabular}{|c|c|c|c|c|}
 \hline
 $q$-range &   ${\scriptstyle v< q < p }$  &  ${\scriptstyle p-\frac{\pi}{2} < q < v  }$  &  ${\scriptstyle v-\frac{\pi}{2} < q < p -\frac{\pi}{2} }$  & ${\scriptstyle   p-\pi <  q < v-\frac{\pi}{2}  }$  \\ 
\hline
 $\nu_{L} $    &  $0$   &   $-\pi\Delta$   &   $-2\pi\Delta$   &   $-2\pi\Delta$   \\ 
 \hline
 $\nu_{R} $  &   $\pi\Delta$ &      $\pi\Delta$     &    $\pi\Delta$     &      $2\pi\Delta$   \\ 
 \hline 
\end{tabular}
\end{center}
\caption{\color{black} $v <  p < u$ }
\label{Tab3}
\end{table}
\setcounter{table}{0}
\renewcommand{\thetable}{\color{blue}{\bf D}}
\begin{table} [http]
\begin{center}
\begin{tabular}{|c|c|c|c|c|}
 \hline
 $q$-range &  ${\scriptstyle p < q < v } $ &  ${\scriptstyle v-\frac{\pi}{2} < q < p } $  &  $ {\scriptstyle p-\frac{\pi}{2} < q < v -\frac{\pi}{2} }  $  &  ${\scriptstyle v-\pi <  q < p-\frac{\pi}{2} } $ \\
\hline
 $\nu_{L} $    &  $-\pi\Delta$   &   $-\pi\Delta$   &   $-\pi\Delta$   &   $-2\pi\Delta$      \\
 \hline
 $\nu_{R} $  &   $0$ &      $\pi\Delta$     &    $2\pi\Delta$     &      $2\pi\Delta$   \\
 \hline 
\end{tabular}
\end{center}
\caption{\color{black} $v-\frac{\pi}{2} <  p <  v$ }
\label{Tab4}
\end{table}

Combining all the above considerations,  we propose the phase assignments for region {\bf I}, {\bf II}, {\bf III} and {\bf IV} to be  made as
\begin{align}    \label{}
\nu_{L} = 
\left\{ \begin{array}{ccc}    
0 &  {\rm for} &  {\bf I} \\
-\pi \Delta  &  {\rm for} &   {\bf II}  \\
-\pi\Delta &  {\rm for} &   {\bf III}    \\
-\pi\Delta & {\rm for} &    {\bf IV} 
\end{array}  \right.   \quad  \,,  \qquad \quad 
\nu_{R} =  \left\{ \begin{array}{ccc}    
\pi\Delta &  {\rm for} &   {\bf I} \\
\pi \Delta  &  {\rm for} &   {\bf II}    \\
 0 & {\rm for}    & {\bf III}   \\
\pi\Delta & {\rm for} &   {\bf IV} 
\end{array}  \right.   \quad \,. 
\end{align}
Thereafter, it is straightforward to assign the appropriate phases on regions ${\bf I}'$, ${\bf II}'$, ${\bf III}'$ and ${\bf IV}'$.  In summary, the phase assignments are given by
\begin{align}    \label{}
\nu_{L} = 
\left\{ \begin{array}{ccc}    
-2\pi\Delta &  {\rm for} &  {\bf I}' \\
-2\pi \Delta  &  {\rm for} &   {\bf II}'  \\
-\pi\Delta &  {\rm for} &   {\bf III}'    \\
-2\pi\Delta & {\rm for} &    {\bf IV}' 
\end{array}  \right.   \quad  \,,  \qquad \quad 
\nu_{R} =  \left\{ \begin{array}{ccc}    
\pi\Delta &  {\rm for} &   {\bf I}' \\
2\pi \Delta  &  {\rm for} &   {\bf II}'    \\
 2\pi\Delta & {\rm for}    & {\bf III}'   \\
2\pi\Delta & {\rm for} &   {\bf IV}'
\end{array}  \right.   \quad \,. 
\end{align}
Finally, one can see that
\begin{equation} \label{phasesum}
\nu_{L}+\nu_{R} = 
\left\{ \begin{array}{ccl}    
\pi\Delta &  {\rm for} &  {\bf I}, {\bf III}' \\
-\pi \Delta  &  {\rm for} &   {\bf III}, {\bf I}'  \\
0 &  {\rm for} &   {\bf II}, {\bf IV}, {\bf II}', {\bf IV}' 
\end{array}  \right.   \,.
\end{equation}
%
 
\section{Some formulae}\label{appb}
\renewcommand{\theequation}{B.\arabic{equation}}
  \setcounter{equation}{0}
In our previous work~\cite{Bak:2018txn},  the dilaton field deformation by the stress-tensor was obtained in the form of 
\begin{align}    \label{DefDil}
\varphi(u,v) & = \int^{u}_{u_{0}}dp~ \frac{\sin(p-u)\cos(p-v)}{\cos(u-v)} T_{uu}(p,v) =\int^{u}_{u_{0}}dp~ \frac{d_{R}(u,v\,;\,p)}{2\cos(u-v)} T_{uu}(p,v)   \,,    \\
T_{uu}(u,v) &= 2\Big[\partial_{u}\partial_{u'}F(u,v|u',v') \Big]_{(u',v')\rightarrow(u,v)}= 2 [ \partial_{u} H_{1}(u,v) -H_{2}(u,v) ] \,,  \nonumber 
\end{align}
where $u_{0}$ is an initial value of $p$ for the non-vanishing $T_{ab}$ and 
\begin{equation} \label{}
H_{1}(u,v)  \equiv  \partial_{u'} F(u,v|u',v') \Big|_{(u',v') \rightarrow (u,v)}\,,  \qquad
 H_{2}(u, v)  \equiv  \partial^{2}_{u'} F(u,v|u',v') \Big|_{(u',v') \rightarrow (u,v)}\,.
\end{equation}
Explicitly, the above function $H_{1}$ could be written as
\begin{equation} \label{}
H_{1}(p,v) = \int^{q_{\tau}}_{q_{i}}dq \Big[ h_{L}(p,v\,;\, q)\, \theta(-d_{R} )   + h_{R}(p,v\,;\, q)\,    \theta(-d_{L} )     \Big]\,, 
\end{equation}
%
%
%
where $d_{L/R} = d_{L/R}(p,v\,;\, q)$ and $h_{L/R}$ are given in Eq.~(\ref{hl}) and (\ref{hr}).  
%

Using the integration by parts, $d_{R}\partial_{p} H_{1} = \partial_{p} (d_{R}H_{1}) - H_{1} \partial_{p}d_{R}$, and recognizing  that $d_{R}(u, v\,;\, u) = d_{R}(u,v\,;\, v-\pi/2)=0$, one can drop the total derivative term   $\partial_{p} (d_{R}H_{1})$  in the integral expression of $\varphi$ (see~\cite{Bak:2018txn}). Apparently, $H_{2}$ contains derivatives of $h_{L/R}$. However, one can rewrite it in terms of $h_{L/R}$ functions without their derivatives. 
By organizing the resultant expression of $\varphi$, one can see, finally, that the deformation of the dilaton field by the double trace interaction can be written as
\begin{align}    \label{Dilaton}
\varphi(u,v)  & =  -2 \int^{u}_{u_{0}}dp \int^{q_{\tau}}_{q_{i}}dq    {\textstyle \bigg[   h_{L} \Big\{ 1+  {\scriptstyle (\Delta+1)}\frac{\sin(u-p)\sin(q-v)}{\cos(u-v)\cos(q-p)}  \Big\}  \theta(-d_{R} )   }  \nonumber \\ 
&\qquad \qquad \qquad \qquad \qquad   {\textstyle  + h_{R} \Big\{ 1- {\scriptstyle (\Delta+1)} \frac{\sin(u-p)\cos(q-v)}{\cos(u-v)\sin(q-p)}    \Big\} \theta(-d_{L} )   \bigg]  }  \,.
\end{align}

To proceed  computing the dilaton deformation, it is useful to perform the  change of variable  from $p$ to $\eta$ for each {\bf I}, {\bf II}, {\bf III} and {\bf IV} region as follows:
\begin{align}    \label{Eta}
\eta(p) &\equiv \left\{ \begin{array}{ll}    
 \frac{\sin(u-p)}{\sin p}\frac{\sin q}{\sin(u-q)}    &  \quad {\rm for} \quad  {\bf I} ~~{\rm and}~~  {\bf II}   \\
   \frac{\cos(v-p)}{\cos p}\frac{\cos q}{\cos(v-q)}  &  \quad {\rm for} \quad  {\bf III}\,,  \\ 
   \frac{\cos(p-q)\sin q}{\sin p}   &  \quad {\rm for} \quad  {\bf IV}  \,. 
\end{array}  \right.   
\end{align}
It would also be useful to introduce $x$ and $y$ variables in each region as
\begin{align}    \label{XandY}
x \equiv  %
\left\{ \begin{array}{ll}    
-\frac{\sin(u-q)}{\sin q}\frac{\cos v}{\cos(u-v)} &  \quad {\rm for} \quad {\bf I} ~{\rm and}~  {\bf II}\,,   \\
\frac{\cos(v-q)}{\cos q}\frac{\cos u}{\cos (u-v)}  & \quad {\rm for} \quad {\bf III}\,,  \\ 
\frac{\cos v}{\sin(q-v) \sin q} & \quad {\rm for} \quad {\bf IV} \,,
\end{array}  \right.   
\qquad y\equiv  
\left\{ \begin{array}{ll}    
-\frac{\sin(u-q)}{\cos(u-q)}\frac{\cos q}{\sin q} & \quad {\rm for} \quad {\bf I} ~{\rm and}~   {\bf II}\,,     \\
-\frac{\cos (v-q)}{\sin (v-q)}\frac{\sin q}{\cos q} &   \quad {\rm for} \quad {\bf III}\,,  \\ 
\frac{\sin u}{\cos (u-q) \sin q}  & \quad {\rm for} \quad {\bf IV}\,.
\end{array}  \right.   
\end{align}
In the  ${\bf I}'$, ${\bf II}'$, ${\bf III}'$ and ${\bf IV}'$ regions, one can introduce $\eta, x, y$ variables just as shifted ones as
\begin{equation} \label{}
\Big[ \eta(p), x, y  \Big] _{{\bf I}', {\bf II}', {\bf III}',  {\bf IV}'} = \Big[\eta(p), x, y \Big]^{q\rightarrow q+\frac{\pi}{2}}_{{\bf I}, {\bf II}, {\bf III}, {\bf IV}}\,.
\end{equation}

Now, let us list some useful formulae  to compute the dilaton field deformation in a closed form,  
First, the integral representation of the Appell's function~\cite{Schlosser:2013hbz} is
\begin{equation} \label{}
\int^{1}_{0}d\eta~   \eta^{a-1}(1-\eta)^{c-a-1}(1-x\eta)^{-b}(1-y\eta)^{-b'}  = B(c-a,a) F_{1}(a,b,b'\,;\, c\,|\, x, y)\,, 
\end{equation}
and its relation to the hypergeometric function is given by
\begin{equation} \label{F1toF}
F_{1}(a,b,b'\,;\,b+b'\,|\,x,y) =  (1-y)^{-a}F\Big(a,b\,;\,b+b'\,|\, z = {\textstyle \frac{x-y}{1-y}  }\Big)\,. 
\end{equation}
Second,  some properties of the hypergeometric function are~\cite{grad} 
\begin{equation} \label{HyperFP}
F(a,b\,;\, c \,|\, z ) = (1-z)^{-b}F(c-a, b\,;\, c\,|\, z)\,,
\end{equation}
\begin{equation} \label{Contig}
c(1-z)F(a,b\,;\, c \,|\, z ) - c F(a-1,b\,;\, c \,|\, z ) + (c-b)zF(a,b\,;\, c+1 \,|\, z )=0\,.
\end{equation}
In  the following, it would also be  useful to introduce new variables $z$ and $w$ as 
\begin{equation} \label{ZandW}
z\equiv \frac{x-y}{1-y}\,, \qquad w\equiv \frac{z}{z-1}\,.
\end{equation}
%

\section{Dilaton field expression }\label{appc}
\renewcommand{\theequation}{C.\arabic{equation}}
  \setcounter{equation}{0}
Here, we present some computational details for the contribution to dilaton deformation in regions {\bf I} and {\bf III}.\\
\underline{Region {\bf I}}:  On the range $v<p<u$ and $v<q<p$, which is denoted as {\bf I} in Fig.~\ref{pqrange}, one can rearrange the integration order as 
\begin{equation} \label{}
\int^{u}_{q_{i}} dp \int^{p}_{q_{i}}dq  = \int^{u}_{q_{i}}dq \int^{q}_{v}dp \,, 
\end{equation}
with the  phase assignments $\nu_{L/R}$ as   
\begin{equation} \label{}
\nu_{L}=0\,, \qquad \nu_{R} = \pi \Delta\,.  \nonumber
\end{equation}
Note also that the  term containing $h_{R}$   in the dilaton expression in Eq.~(\ref{DilatonM})   vanishes because of  $d_{L} >0$. Now, let us present  computational steps to arrive at the result in Eq.~(\ref{varphiRI}). 
First,  Eq.~(\ref{hl}) under the above conditions leads to  the relevant function $h_{L}$  in the form of 
\begin{align}    \label{}
h_{L}(p,v\,;\,q)&= -  { \textstyle 2\Delta Q_{\Delta} \frac{\cos^{2\Delta-1}(p-v)}{\cos^{\Delta+1}(q-p)\sin^{\Delta-1}(q-v)} \frac{1}{\sin^{\Delta}(p-q)\cos^{\Delta}(q-v)} }\,,
\end{align}
where $Q_{\Delta}$ is defined by $Q_{\Delta} \equiv \frac{\bar{h} \bar{N}_{\Delta} }{2^{2\Delta}} \sin\pi\Delta=\frac{\bar{h} N_{\Delta}  }{2^{2\Delta}} $.
By the change of variable  in Eq.~(\ref{Eta}) 
and by using  the integral representation of the Appell's function,
one can see that 
\begin{align}    \label{hlint1}
\int^{q}_{v}dp~ h_{L} &= 2\Delta Q_{\Delta}\int^{1}_{0}d\eta~   {\textstyle  \frac{\sin^{2}p}{\sin u}\frac{\sin(u-p)}{\sin q} } h_{L}  \nonumber \\
&={\textstyle  2Q_{\Delta}\frac{\sin u}{\cos q} \frac{\cos^{2\Delta-1}(u-v)}{\cos^{\Delta}(u-q)\sin^{\Delta}(u-q)} \frac{1}{\sin^{\Delta-1}(q-v)\cos^{\Delta}(q-v)}  }  F_{1}(1\,;\,1-2\Delta,\Delta\,;\,2-\Delta\,|\,x,y)\,,
\end{align}
where  $x$ and $y$ are given in Eq.~(\ref{XandY}).  Using the relation of Appell's function to the hypergeometric function in Eq.~(\ref{F1toF}) and the property of the hypergeometric function in Eq.~(\ref{HyperFP}), one can also see that the above expression reduces to a  hypergeometric function 
\begin{equation} \label{}
\int^{q}_{v}dp~ h_{L} =  2Q_{\Delta}\,  w^{-\Delta} F(-\Delta,1-2\Delta\,;\, 1-\Delta\,|\, w)\,,
\end{equation}
where $w$ is given in Eq.~(\ref{ZandW}) and becomes in this region
\begin{equation} \label{}
w(u,v\,;\, q) = \frac{\sin (u-q)}{\cos(u-q)}\frac{\sin (q-v)}{\cos(q-v)}\,.
\end{equation}
Repeating a similar computation for the remaining term in Eq.~(\ref{Dilaton}), one obtains
\begin{align}    \label{hlint2}
& \int^{q}_{v}dp~   {\textstyle h_{L}  \bigg[     {\scriptstyle (\Delta+1)}\frac{\sin(u-p)\sin(q-v)}{\cos(u-v)\cos(q-p)}   \bigg] }   \nonumber \\
&=  - {\textstyle  2\Delta Q_{\Delta}\frac{\Gamma(-\Delta)}{\Gamma(1-\Delta)}  \frac{\sin^{2} u}{\cos^{2} q} \frac{\cos^{2\Delta-2}(u-v)}{\cos^{\Delta}(u-q)\sin^{\Delta}(u-q)} \frac{1}{\sin^{\Delta-2}(q-v)\cos^{\Delta}(q-v)}  }  F_{1}(1\,;\,1-2\Delta,\Delta\,;\,2-\Delta\,|\,x,y)\,,\nonumber \\
&=  - 2Q_{\Delta}\frac{w^{-\Delta}}{1-w} F(-1-\Delta,1-2\Delta\,;\, 1-\Delta\,|\, w)\,.
\end{align}

Combining the results in Eq.~(\ref{hlint1}) and~(\ref{hlint2}) and then using the property of the hypergeometric function in Eq.~(\ref{Contig}), one can see that the contribution to the dilaton field deformation from the region {\bf I} is given by
\begin{equation} \label{}
\varphi_{\bf I} (u,v\,;\, q_{i})= 4\Delta Q_{\Delta} \int^{u}_{q_{i}} dq \Big[w^{1-\Delta}(1-w)^{2\Delta-1} + \Delta \frac{1+w}{1-w}B_{w}(1-\Delta,2\Delta)\Big]\,,
\end{equation}
where  $B_{w}$ denotes the incomplete Beta function 
\begin{equation} \label{}
B_{w}(a,b)\equiv \int^{w}_{0}dt~ t^{a-1}(1-t)^{b-1} = \frac{w^{a}}{a}F(a,1-b,;\, 1+a\,|\, w)\,. \nonumber 
\end{equation}
As in Ref.~\cite{Bak:2018txn}, the asymptotic expansion of the above expression around $w=1$ can be obtained in the form of 
\begin{equation} \label{asympphi1}
\varphi_{\bf I} (u,v\,;\, q_{i})= \frac{\bar{h}}{4\pi}  \Delta B(\Delta,\Delta)\Big[ \frac{\sin \mu + \sin( \tau -2q_{i})  }{\cos \mu} \Big]+ \cdots\,.
\end{equation}

When $q_{i}=v$,  the integral expression of the dilaton field deformation $\varphi_{\bf I}$ could be further simplified. In this case, one may note that $q$-integration could be rewritten in terms of $w$ as

\begin{equation} \label{}
\int^{u}_{v}dq = \int^{\tan^{2}\frac{u-v}{2}}_{0}\frac{dw}{(1-w)^{2}} {\textstyle \frac{2}{\sqrt{\frac{1}{ \cos^{2}(u-v)} -(\frac{1+w}{1-w})^{2}}}  }\,. 
\end{equation}
Using the integration by parts,
\begin{align}    \label{}
& {\textstyle \frac{2}{(1-w)^{2}}\frac{1}{\sqrt{\frac{1}{\cos^{2}(u-v)}-(\frac{1+w}{1-w})^{2} } } \Delta \frac{1+w}{1-w}B_{w}(1-\Delta,2\Delta) }  \nonumber \\
&=  -  {\textstyle \frac{d}{dw}\Big[ \sqrt{\frac{1}{\cos^{2}(u-v)}-(\frac{1+w}{1-w})^{2}}\Delta  B_{w}(1-\Delta,2\Delta)  \Big]  + \sqrt{\frac{1}{\cos^{2}(u-v)}-(\frac{1+w}{1-w})^{2}}\, \Delta w^{-\Delta}(1-w)^{2\Delta -1} }\,,  \nonumber 
\end{align}
and noting that the total derivative term vanishes at  the limit points of integral  {\it i.e.} $w=0$ and $w=\tan^{2}\frac{u-v}{2}$, 
one can see that  the expression of $\varphi_{\bf I}=\varphi_{\bf I}(u,v) $  reduces to
\begin{equation} \label{}
\varphi_{\bf I} = 4\Delta Q_{\Delta} \int^{\tan^{2}\frac{u-v}{2}}_{0} dw  \bigg[{\textstyle \sqrt{\frac{1}{\cos^{2}(u-v)}-(\frac{1+w}{1-w})^{2}} }~\Delta  w^{-\Delta}(1-w)^{2\Delta-1} + \frac{2w^{1-\Delta}(1-w)^{2\Delta-3}}{\sqrt{\frac{1}{\cos^{2}(u-v)}-(\frac{1+w}{1-w})^{2}}}\bigg]\,.
\end{equation}
Now, it is useful to perform a change of variable from $w$ to $\zeta$ as 
\begin{equation} \label{}
1 +\tan^{2}(u-v)\, \zeta(w) \equiv \Big(\frac{1+w}{1-w}\Big)^{2}\,,
\end{equation}
which leads to  
\begin{equation} \label{}
\int^{\tan^{2}\frac{u-v}{2}}_{0}\frac{dw}{(1-w)^{2}}  =  \int^{1}_{0}d\zeta \frac{\tan^{2}(u-v)}{4\sqrt{1+ \zeta \tan^{2}(u-v) }}\,.
\end{equation}
Finally,  by using the integral representation of hypergeometric function and using its transformation property~\cite{grad}, one obtains the closed form of the dilation deformation in the region {\bf I} as
\begin{align}    \label{}
\varphi_{\bf I} & = {\textstyle 2\Delta Q_{\Delta}   B(1-\Delta,1-\Delta)  \frac{\tan(u-v)}{\sin^{2\Delta}(u-v)} \Big[ F \big(1-
\Delta,-\Delta\,;\,\frac{3}{2}-\Delta\, |\, {\scriptstyle \sin^{2}(u-v) } \big)  } \nonumber \\
& \qquad \qquad \qquad\qquad \qquad \qquad   -  {\textstyle  {\scriptstyle \cos^{2\Delta+1}(u-v)} F \big(\frac{1}{2},\frac{1}{2}\,;\,\frac{3}{2}-\Delta\, |\, {\scriptstyle \sin^{2}(u-v) } \big) \Big] }  \nonumber  \\
&= {\textstyle 4\Delta Q_{\Delta}\frac{B(2-\Delta,2-\Delta)}{1-\Delta}{\scriptstyle \tan(u-v)\sin^{2-2\Delta}(u-v) } F\big(1-\Delta,1-\Delta\,;\, \frac{5}{2}-\Delta\,|\, {\scriptstyle \sin^{2}(u-v) } \big) }
 \,.
\end{align}

Near the right boundary, one may set $u-v = \frac{\pi}{2} - \xi$ with $0 < \xi \ll 1$. By using the following transformation property of the hypergeometric function  
\begin{align}    \label{}
F(a,b\,;\, c\, |\, z) = & \frac{\Gamma(c)\Gamma(c-a-b)}{\Gamma(c-a)\Gamma(c-b)}F(a,b\,;\, a+b-c+1\,|\, 1-z) \\
 & \qquad + (1-z)^{c-a-b}\frac{\Gamma(c)\Gamma(a+b-c)}{\Gamma(a)\Gamma(b)} F(c-a,c-b\,;\, c-a-b+1\, | \, 1-z)\,,\nonumber 
\end{align}
one obtains near the right boundary 
\begin{align}    \label{}
 \label{asymp}
\varphi_{\bf I} & = 2\Delta  Q_{\Delta} B(1-\Delta,1-\Delta) {\textstyle \frac{\Gamma(\frac{3}{2}-\Delta) \Gamma(\frac{1}{2}+\Delta)}{\Gamma(\frac{1}{2})\Gamma(\frac{3}{2})}  } \frac{1}{\sin\xi} + {\cal O}(\sin^{2\Delta}\xi ) \nonumber \\
&= \frac{\bar{h}\Delta^{2} N_{\Delta}}{2^{2\Delta-2}} B(1-\Delta,2\Delta) \frac{1}{\sin\xi} + {\cal O}(\sin^{2\Delta}\xi ) \nonumber \\
&= \frac{\bar{h}}{2\pi}\Delta B(\Delta,\Delta) \frac{1}{\sin\xi} + {\cal O}(\sin^{2\Delta}\xi ) \,. 
\end{align}

\noindent \underline{Region {\bf III}}:
Now, let us consider the contribution from the region {\bf III}. From the signature of $d_{L/R}$ and the phase assignment as
\begin{equation} \label{hrint1}
\nu_{L} =  -\pi \Delta\,, \qquad \nu_{R} = 0\,,
\end{equation}
one can see that  Eq.~(\ref{hr}) leads to 
\begin{equation} \label{}
h_{R}(p,v\,;\, q) = -2\Delta Q_{\Delta}\frac{\cos^{2\Delta-1}(p-v)}{\cos^{\Delta}(q-p)\sin^{\Delta}(v-q)} \frac{1}{\sin^{\Delta+1}(q-p)\cos^{\Delta-1}(q-v)}\,,
\end{equation}
where $Q_{\Delta}$ was introduced before as $Q_{\Delta} = \frac{\bar{h} N_{\Delta} }{2^{2\Delta}}$. Note that the overall minus sign comes from $\sin(\nu_{L}+\nu_{R}) = - \sin \pi \Delta$ in this region. 
Following the same change of variables in Eq.~(\ref{Eta}) and~(\ref{XandY}) and using $\eta$ and  $x, y$ variables, one can also see that 
\begin{align}    \label{hrint2}
& \int^{q}_{v-\frac{\pi}{2}}dp~ {\textstyle   h_{R} \Big\{ 1- {\scriptstyle (\Delta+1)} \frac{\sin(u-p)\cos(q-v)}{\cos(u-v)\sin(q-p)}    \Big\}  }  \nonumber \\
 &= -2\Delta Q_{\Delta}\int^{1}_{0}d\eta~   {\textstyle  \frac{\cos(v-q)}{\cos q}\frac{\cos^{2} p}{\sin v} } {\textstyle   h_{R} \Big\{ 1- {\scriptstyle (\Delta+1)} \frac{\sin(u-p)\cos(q-v)}{\cos(u-v)\sin(q-p)}    \Big\}  }   \nonumber \\
&= -2\Delta Q_{\Delta}  {\textstyle \Big[\frac{\sin v}{\cos q\sin(v-q)}\Big]^{2\Delta}   B(-\Delta,2\Delta)}  \Big[ F_{1}(2\Delta\,;\,0,\Delta\,;\,\Delta\,|\,x,y)   \nonumber \\
& \hskip5cm  \qquad \qquad \qquad   + (\Delta -1) F_{1}(2\Delta\,;\,-1,\Delta\,;\,\Delta-1\,|\,x,y)  \Big]\,, \nonumber \\
&= - 2\Delta^{2}Q_{\Delta}   B(-\Delta,2\Delta) (1-2z)\,,
\end{align}
where we used Eq.~(\ref{F1toF})  and Eq.~(\ref{HyperFP}). Here,  $z$ is defined in Eq.~(\ref{ZandW}) and is given in this region by
\begin{equation} \label{}
z = \frac{1}{2}\Big[1+\frac{\cos(u+v-2q)}{\cos(u-v)}\Big]\,. \nonumber
\end{equation}
Combining results in Eq.~(\ref{hrint1}) and~(\ref{hrint2}), one can see that the contribution to the dilaton deformation is given by 
\begin{align}    \label{}
\varphi_{\bf III} (u,v\,;\, q_{i}) &= 4\Delta^{2} Q_{\Delta} B(-\Delta,2\Delta)  \int^{v}_{q_{i}}  dq \frac{\cos(u+v-2q)}{\cos(u-v)} \nonumber \\
&= 2\Delta^{2} Q_{\Delta}B(-\Delta,2\Delta)\Big[ \tan (u-v)  - \frac{\sin(u+v-2q_{i})}{\cos(u-v)}\Big]\,. 
\end{align}
When $q_{i} $ takes its lowest value $v-\frac{\pi}{2}$, the dilaton deformation is further simplified to 
\begin{equation} \label{}
\varphi_{\bf III} = 4\Delta^{2} Q_{\Delta}B(-\Delta,2\Delta) \tan \mu\,. 
\end{equation}

\end{document}